\documentclass[aps,pre,preprint,showpacs,floatfix,unsortedaddress,
superscriptaddress,nofootinbib]{revtex4-1}

\usepackage{epsfig}
\usepackage{float}
\usepackage{latexsym}
\usepackage{amsmath}
\usepackage{amsfonts}
\usepackage{amssymb}
\usepackage{amsbsy}
\usepackage{subfigure}
\usepackage{bm}
\usepackage{color}
\usepackage{dcolumn}
\usepackage{hyperref}
\usepackage{graphicx}
\usepackage{subeqnarray}
\usepackage{bm}
\usepackage[normalem]{ulem}

\usepackage{silence}
\graphicspath{{./figures/}}

\begin{document}
\title{Corner transport upwind lattice Boltzmann model for bubble cavitation}
\author{V. Sofonea}
\email[]{sofonea@gmail.com}
\affiliation{Center for Fundamental and Advanced Technical Research, 
Romanian Academy, Bd. Mihai Viteazul 24, 300223 Timi\c{s}oara,
Romania}
\author{T. Biciu\c{s}c\u{a}}
\email[]{biciusca.tonino@gmail.com}
\affiliation{Center for Fundamental and Advanced Technical Research, 
Romanian Academy, Bd. Mihai Viteazul 24, 300223 Timi\c{s}oara, Romania}
\affiliation{Department of Physics, West University of Timi\c{s}oara,
Bd. Vasile P\^{a}rvan 4, 300223 Timi\c{s}oara, Romania}
\author{S. Busuioc}
\email[]{sergiu.busuioc@e-uvt.ro}
\affiliation{Center for Fundamental and Advanced Technical Research, 
Romanian Academy, Bd. Mihai Viteazul 24, 300223 Timi\c{s}oara, Romania}
\affiliation{Department of Physics, West University of Timi\c{s}oara,
Bd. Vasile P\^{a}rvan 4, 300223 Timi\c{s}oara, Romania}
\author{Victor E. Ambru\c{s}}
\email[]{victor.ambrus@e-uvt.ro}
\affiliation{Center for Fundamental and Advanced Technical Research, 
Romanian Academy, Bd. Mihai Viteazul 24, 300223 Timi\c{s}oara, Romania}
\affiliation{Department of Physics, West University of Timi\c{s}oara,
Bd. Vasile P\^{a}rvan 4, 300223 Timi\c{s}oara, Romania}
\author{G. Gonnella}
\email[]{gonnella@ba.infn.it}
\affiliation{Dipartimento di
Fisica, Universit\`{a} di Bari,
 {\it and} INFN, Sezione di Bari,
Via Amendola 173, 70126 Bari, Italy}
\author{A. Lamura}
\email[]{Corresponding author, a.lamura@ba.iac.cnr.it}
\affiliation{
Istituto Applicazioni Calcolo, CNR,
Via Amendola 122/D, 70126 Bari, Italy}

\date{\today}
\begin{abstract}
Aiming to study the bubble cavitation problem in quiescent and sheared liquids, 
a third-order isothermal lattice Boltzmann (LB) model that describes
a two-dimensional ($2D$) fluid obeying the
van der Waals equation of state, is introduced. The evolution 
equations for the distribution functions in this off-lattice model with
16 velocities are solved
using the corner transport upwind (CTU) numerical scheme on large
square lattices (up to $6144 \times 6144$ nodes). 
The numerical viscosity and the regularization of the model are discussed
for first and second order CTU schemes finding that the latter choice
allows to obtain a very accurate phase diagram of a nonideal fluid.
In a quiescent liquid,
the present model allows to recover the solution of the
$2D$ Rayleigh-Plesset equation for a growing vapor bubble. 
In a sheared liquid, we
investigated the evolution of the total bubble area, the bubble deformation
and the bubble tilt angle, for various values of the shear rate.  
A linear relation between the dimensionless deformation 
coefficient $D$  and the capillary number $Ca$ is found at small $Ca$ 
but with a different factor than in equilibrium liquids.
A non-linear regime is observed for $Ca \gtrsim 0.2$. 
\end{abstract}
\maketitle

\section{Introduction}

The cavitation and growth of bubbles in stretched 
or superheated liquids is a phenomenon
frequently appearing in nature with relevant 
scientific and technical interest \cite{rallison-1984}. 
Cavitation is a sudden transition from liquid to vapor that can be
promoted by the decrease of the pressure in a stretched liquid  below
the liquid's vapor pressure as well as by the nucleation of bubbles
in a superheated liquid \cite{brennen-1995}. Examples of these 
processes, among others,  are given by the cavitation corrosion of materials
exposed to water \cite{plesset-1963}, phase changes in 
cosmology \cite{devega-2001}, vulcanism \cite{massol-2005}.
In the following we will be interested in studying numerically 
the kinetics and dynamics of a single vapor bubble which cavitates
in a superheated liquid which is either at rest or subject to shear.
Previous studies of a nucleating bubble are very limited and rely
on Molecular Dynamics \cite{kuksin-2010,watanabe-2010,diemand-2014,
angelil-2014}, lattice Boltzmann (LB) simulations 
\cite{sukop-2005,chen-2010,chen-2011,zhong-2012}, and other numerical methods
\cite{feng-1997}.
Growth curves of the bubble in a quiescent fluid were obtained in 
Refs.~\cite{angelil-2014,chen-2010,chen-2011} and compared to the
Rayleigh-Plesset (RP) 
growth model \cite{rayleigh-1917,plesset-1949,plesset-1977}.
Very first attempts of addressing the cavitation study in a sheared liquid
were presented in Refs.~\cite{chen-2010,chen-2011}.

From more than two decades, the use of LB models for phase-separating
fluids is widely expanding because of the parallel nature of their basic
algorithm, as well as for their capability to easily handle 
interactions \cite{chen_annurev1998,succi_2001,sukop_2006,aidun_annurev2010,guo2013,sukop2015,krueger2017}.
A characteristic feature of the LB models is the polynomial
expansion of the equilibrium single-particle 
distribution function up to a certain
order $N$ with respect to the fluid particle velocity. This expansion is made
by projecting the equilibrium distribution function on a set of orthogonal
polynomials, e.g., the Hermite polynomials \cite{shan_jfm2006}.
In the widely used {\emph{collision-streaming}} LB
models, the velocity space is discretized so that the velocity vectors
of the fluid particles leaving a node of the lattice are oriented towards the
neighboring nodes \cite{cao1997}. Such models are
also called {\emph{on-lattice}} models. 
 
In this paper we perform a qualitative and quantitative analysis
of the bubble cavitation problem using
a {\emph{third-order}} isothermal LB model that describes 
a two-dimensional (2D) nonideal fluid obeying the
van der Waals equation of state (EOS) \cite{dsfd2014}. 
Though several equations of state exist 
\cite{yuan2006}
and different lattice Boltzmann models are available to handle 
high liquid-vapor density ratios \cite{chen2014}, 
the used EOS
is a well-established and classic benchmark
fitting our goal. 
Indeed, a recent numerical study
\cite{kaehler-2015}, based on the van der Waals EOS, 
allowed to elucidate qualitatively and quantitatively the
cavitation inception at a sack-wall obstacle in a 2D geometry.
The study of 
two-dimensional bubbles has attracted a lot of interest in the past.
Indeed, an immiscible drop in shear flow has been studied theoretically
\cite{rich1968,buck1973}
and numerically \cite{hall1996,hall21996}. 
For two-dimensional
miscible binary mixtures the problem of bubble break-up and dissolution 
under shear was also addressed \cite{wagner1997}. 

The $2D$ LB model used in this paper, which is described in
Sections \ref{sec:model}\,A-C, has 16 
\emph{off-lattice} velocities and is based on the Gauss-Hermite
quadrature method \cite{dsfd2014,shan_jfm2006}. 
In Ref.\cite{dsfd2014}, the evolution equations for the distribution functions
in the LB model were solved using the first order
corner transport upwind (CTU1) numerical scheme 
\cite{dsfd2014,colella1990,leveque1996,leveque2002,trangenstein2009}.
Besides the capability of handling off-lattice velocity sets in LB models,
this very simple scheme, which is of first order with
respect to the lattice spacing $\delta s$,
involves only four neighboring lattice nodes and
is easily parallelizable, like the collision-streaming scheme. 
Despite of these advantages, the computer simulations performed
with the CTU1 scheme are plagued by its numerical viscosity, as discussed
in Section \ref{nvsection} below. To improve
the accuracy of our simulations, in this paper we further extended the
previous LB model \cite{dsfd2014}
by incorporating the second-order corner transport upwind scheme
(CTU2) \cite{leveque1996,leveque2002,trangenstein2009}.
These schemes, though well documented in the
mathematical literature for the numerical solution
of hyperbolic partial differential equations,
are here demonstrated to have the capabilities 
to deal with an off-lattice discrete velocity set in a 
LB model, and the provided results are encouraging.

In order to follow the bubble evolution on large lattices during long
time intervals, we implemented this model on  NVIDIA$^{\textregistered}$ 
graphics processing units (M2090 and K40). The resulted code was
first tested  by simulating the evolution of shear waves oriented along
the horizontal axis or along the diagonal of a square lattice. During these
simulations, we checked for anisotropic effects in the LB model and we
found that no regularization procedure is needed
for small values of the relaxation time ($\tau \leq 0.1$), i.e., when the
isothermal fluid is not too far from equilibrium and obeys the mass
and momentum conservation equations (Section \ref{nvsection}).
Further tests reported in Section \ref{nvsection2}
refer to the liquid-vapor phase diagram and to the effect of both the relaxation
time $\tau$ and the lattice spacing $\delta s$ on the accuracy of the
liquid and vapor density values obtained by equilibrating a plane
interface.

Since the growth or shrinkage of a bubble mainly depends on its initial size
at fixed temperature and pressure, in Section \ref{snuma}
we checked the theoretical prediction
\cite{laurila-2012} of the critical radius of the bubble neither growing
nor shrinking in a quiescent superheated liquid. 
In such a system the bubble Helmholtz free energy density can decrease by
increasing the bubble size via evaporation
of some of the surrounding liquid to the
coexistence densities. Alternatively,
the interfacial free energy increases as the bubble
shrinks. 
The competition between
these two mechanisms, under the constraint of local mass conservation,
induces either the growth or the collapse of the bubble.

When the bubble cavitates,
the time evolution of its radius can be theoretically described
by the RP model \cite{rayleigh-1917,plesset-1949,plesset-1977}, where
the Navier-Stokes equation is re-written for a spherical bubble in an infinite
liquid domain. In Section \ref{srp} of this paper
 we derive the RP equation in two dimensions and compare
our numerical findings to its predictions.
This will allow to test the accuracy of the present off-lattice
numerical model in addressing the problem of cavitation.
Indeed, the RP equation is useful 
to quantitatively characterize the growth of bubbles
in cavitation. This problem is often tackled in two dimensions
due to its heavy computational cost \cite{falcucci-2013,kaehler-2015}. 
In this way the analysis of the RP equation
in a low dimensionality system may give an analytical support
to further numerical studies. Our study shows that the numerical model
gives the right growth rate of a cavitating bubble 
up to a final bubble size which is more than one order of magnitude larger
than its initial value.

Finally, despite the deep scientific and technological interest for the problem
of the deformation of a bubble in an immiscible fluid under an external flow
\cite{rallison-1984}, the growth of a vapor bubble in shear flow
has not been the 
subject of extended investigation. In the present study we are able
to characterize the growth and the deformation of the bubble on 
time scales long enough to access non-negligible values of the capillary
number (Section \ref{sshear}). 
Moreover, the tilt angle of the deformed bubble
with respect to the flow direction and its areal extension are computed.

In this paper, all physical quantities are nondimensionalized by using the
following reference quantities \cite{pre2004}:
the fluid particle number density $n_{R} = N_{A} / V_{mc}$, the critical
temperature $T_{R} = T_{c}$, the fluid particle mass
$m_{R} = M / N_A$, the length $l_{R} = 1/\sqrt[3]{n_{R}}$,
the speed $c_{R} = \sqrt{k_{B}T_{R}/m_{R}}$,
and the time $t_{R} = l_{R}/c_{R}$. Here
$N_{A}$ is  Avogadro's number, $V_{mc}$ is the molar volume at the critical point, $T_{c}$ is the critical temperature and
$M$ is the molar mass.


\section{Description of the model} \label{sec:model}


\subsection
{Velocity set, single-particle distribution functions and evolution equations}


In order to derive the Navier-Stokes equations
from the Boltzmann equation in the case of
a compressible isothermal fluid 
\cite{shan_jfm2006,ambrus_pre2012,ambrus_jcph2016},
the moments up to the order $N=3$ of the Maxwell - Boltzmann equilibrium
single-particle distribution function
\begin{equation}
f^{eq} \equiv f^{eq}({\bm{x}},{\bm{\xi}},t) =
\frac{\rho}{\,(2\pi T)^{D/2}\,}
\exp\left[-\frac{\,\,({\bm{\xi}} - {\bm{u}})^{2}\,}{2T}\right]
\label{efeq}
\end{equation}
are required 
according to the Chapman-Enskog method
\cite{chen_annurev1998,succi_2001,sukop_2006,aidun_annurev2010,guo2013,sukop2015,krueger2017}.
In Eq.~\eqref{efeq} above, ${\bm{x}}$ is the fluid particle position vector,
${\bm{\xi}}$ is the fluid particle velocity vector, $t$ is the time and
$\rho\equiv\rho({\bm{x}},t)$, $T\equiv T({\bm{x}},t)$,
${\bm{u}}\equiv {\bm{u}}({\bm{x}},t)$ are the local values of the
fluid 
particle number density, fluid temperature and fluid velocity, respectively.
In the {\emph{Gauss - Hermite}} LB model of order $N$
in $D$ dimensions (see \cite{shan_jfm2006} and references therein), 
the equilibrium single-particle distribution function \eqref{efeq}
is expanded up to order
$N$ with respect to the tensor Hermite polynomials
${\bm{\mathcal{H}}}^{(\ell)}({\bm{\xi}}) \equiv
{\bm{\mathcal{H}}}^{(\ell)}_{\alpha_{1} \ldots \alpha_{\ell}}({\bm{\xi}})$,\,
$0\leq \ell \leq N$ ($1 \leq \alpha_{1},\,\ldots \alpha_{l} \leq D$) :
\begin{equation}
f^{eq} ({\bm{x}},{\bm{\xi}},t) \, = \,
 \omega({\bm{\xi}}) \, \sum_{\ell=0}^{N} \,
\frac{1}{\ell!}  \, {\bm{a}}^{eq,(\ell)}_{\alpha_{1} \ldots \alpha_{\ell}} 
({\bm{x}},t)\,
{\bm{\mathcal{H}}}^{(\ell)}_{\alpha_{1} \ldots \alpha_{\ell}}({\bm{\xi}})
\label{exfeqh} 
\end{equation}
where summation over
repeated lower Greek indices is implicitly
understood and 
\begin{eqnarray}
\omega({\bm{\xi}}) & = & \frac{1}{\,2\pi\,}\,e^{-{\bm{\xi}}^{2}/2T} \nonumber \\
{\bm{a}}^{eq,(\ell)}_{\alpha_{1} \ldots \alpha_{\ell}} 
({\bm{x}},t) & = & \int f^{eq} ({\bm{x}},{\bm{\xi}},t) 
{\bm{\mathcal{H}}}^{(\ell)}_{\alpha_{1} \ldots \alpha_{\ell}}({\bm{\xi}})
d{\bm{\xi}} .\rule{0mm}{8mm} \label{exfeqa}
\end{eqnarray}
All the moments up to order $N$ of $f^{eq}({\bm{x}},{\bm{\xi}},t)$, namely
$\int f^{eq}({\bm{x}},{\bm{\xi}},t) \xi_{\alpha_{1}}
\ldots \xi_{\alpha_{N}} d{\bm{\xi}}$,
are thereafter
recovered using appropriate quadrature methods in the velocity space
\cite{shan_jfm2006,ambrus_pre2012,ambrus_jcph2016,shan_prl1998,piaud_ijmpc2014}.

The Gauss-Hermite quadrature method
\cite{shan_jfm2006,hildebrandt,shizgal} allows one to get
a finite set of velocity vectors  (quadrature points) ${\bm{\xi}}_{k}$,
$k=1,\,2,\,\ldots\,K$, as well as their associated weights $w_{k}$.
The expansion \eqref{exfeqh}, followed by the application of the
Gauss-Hermite quadrature method leads to the LB model, where
the Boltzmann equation is replaced by a set of evolution equations
for the functions $f_{k}\equiv f_{k}({\bm{x}},t)=f({\bm{x}},{\bm{\xi}}_{k},t)$,
which are usually defined in the nodes ${\bm{x}}$ of a regular lattice.
When using the BGK collision term in a $D$-dimensional LB model
of order $N$
\cite{shan_jfm2006,chen_annurev1998,succi_2001,sukop_2006,ambrus_jcph2016,shan2008},
the functions $f_{k}$, $1\leq k \leq K = (N+1)^{D}$, evolve according to
\begin{equation}
 \partial_{t} f_{k} \,+\, \xi_{k,\gamma} \,\partial_{\gamma} f_{k} \, = \,
        -\,\frac{1}{\,\tau\,}\,\left[\, f_{k}\,-\,f^{eq}_{k}\,\right]\,+\, F_{k}
         \qquad , \qquad
        1 \leq k \leq K
\label{eBGK}
\end{equation}
where
$\partial_{t}=\partial / \partial_{t}$,\,
$\xi_{k,\gamma}$, $\gamma \in \{x,y,\ldots\}$,
are the Cartesian components of the velocity vector
${\bm{\xi}}_{k}$, $\partial_{\gamma}=\partial / \partial_{x_\gamma}$,
\begin{equation}
f_{k}^{eq} \equiv f_{k}^{eq}({\bm{x}},t) \, = \,
 w_{k} \, \sum_{\ell=0}^{N} \,
\frac{1}{\ell!}  \, {\bm{a}}^{eq,(\ell)}_{\alpha_{1} \ldots \alpha_{\ell}} 
({\bm{x}},t)\,
{\bm{\mathcal{H}}}^{(\ell)}_{\alpha_{1} \ldots \alpha_{\ell}}({\bm{\xi}}_{k}) ,
\label{exfeqh2} 
\end{equation}
and $\tau$ is the relaxation time. In the Gauss - Hermite LB model
of order $N=3$,  the expressions of the
functions $f_{k}^{eq}\equiv f_{k}^{eq}({\bm{x}},t)$
and of the force term $F_{k}$ are
\cite{shan_jfm2006,niu2007pre,suga2010pre,suga2013fdr} :
\begin{eqnarray}
f^{eq}_{k} & = & w_{k}\rho \left\{\,1\,+\,{\bm{\xi}}_{k}\cdot {\bm{u}} \,+\,
\frac{1}{\,2\,} \left[ ({\bm{\xi}}_{k}\cdot {\bm{u}})^{2} \,-\, u^{2}\,+\,
(T-1)(\bm{\xi}^{2}_{k}-2) \right] \right. \nonumber\\
& + & \left. \frac{\, {\bm{\xi}}_{k}\cdot {\bm{u}}  \,}{6} \, 
\left[ ({\bm{\xi}}_{k}\cdot {\bm{u}})^{2}
- 3 u^{2} \,+\, 3(T-1)(\bm{\xi}^{2}_{k}-4) \right] \right\}
\label{herfeq} \rule{0mm}{9mm} \\
F_{k} & = & w_{k}\rho \left\{\,{\bm{\xi}}_{k}\cdot {\bm{g}} \,+\,
({\bm{\xi}}_{k}\cdot {\bm{g}}) ({\bm{\xi}}_{k}\cdot {\bm{u}})
\,-\, {\bm{g}}\cdot {\bm{u}}
\,+\, \frac{1}{\,2\rho\,} {\bm{a}}^{(2)}
 \left[ \,({\bm{\xi}}_{k}\cdot {\bm{g}})
{\bm{\mathcal{H}}}^{(2)}({\bm{\xi}}_{k})
 \,-\, 2{\bm{g}}{\bm{\xi}}_{k}
 \right] \right\}
 \label{herforce}
\end{eqnarray}
where
\begin{eqnarray}
 \rho\equiv \rho({\bm{x}},t) & = & \sum_{k=1}^{K} f_{k} \,=\,
 \sum_{k=1}^{K} f_{k}^{eq}
 \\
{\bm{u}}\equiv {\bm{u}}({\bm{x}},t) & = &
\frac{1}{\,\rho\,} \sum_{k=1}^{K} f_{k} {\bm{\xi}}_{k} \,=\,
\frac{1}{\,\rho\,} \sum_{k=1}^{K} f_{k}^{eq} {\bm{\xi}}_{k}
\rule{0mm}{9mm}  
\end{eqnarray}
are the local density and velocity.
In the expression \eqref{herforce} of $F_{k}$,
${\bm{g}}$ is an acceleration depending
on the specific problem that is investigated with the LB model.
For the model used in this paper, ${\bm{g}}$  is given in
Eq.~(\ref{eder}) below.

\begin{table}
\caption{The Cartesian projections of the vectors ${\bm{\xi}}_{k}$,\,
$k = 1,\,2,\,\ldots K=16$, and their corresponding weights
$w_{k}$  used in the
two-dimensional isothermal LB model of order $N=3$  \cite{shan_jfm2006,dsfd2014}.}
\label{roots}
\begin{center}{\vbox{\vspace{5mm}}}
\begin{tabular}{|c|ccc|}
\hline
\quad $k$ \quad
 & $\xi_{k,x}$ & $\xi_{k,y}$ & $w_{k}$  \\  \hline\hline
\rule{0mm}{8mm} 1 \ldots 4 &
\quad $\pm \sqrt{3-\sqrt{6}}$ \quad &
 \quad $\pm \sqrt{3-\sqrt{6}}$ \quad &
 \quad $(5 + 2\sqrt{6}) / 48$ \qquad \\

\rule{0mm}{6mm} 5 \ldots 8 &
\quad $\pm \sqrt{3+\sqrt{6}}$ \quad &
 \quad $\pm \sqrt{3-\sqrt{6}}$ \quad &
 \quad $1 / 48$ \quad \\

\rule{0mm}{6mm} 9 \ldots 12 &
\quad $\pm \sqrt{3-\sqrt{6}}$ \quad &
 \quad $\pm \sqrt{3+\sqrt{6}}$ \quad &
 \quad $1 / 48$ \quad \\

\rule{0mm}{6mm} 13 \ldots 16 &
\quad $\pm \sqrt{3+\sqrt{6}}$ \quad &
 \quad $\pm \sqrt{3+\sqrt{6}}$ \quad &
 \quad $(5 - 2\sqrt{6}) / 48$ \quad \\
[1ex] \hline
\end{tabular}
\end{center}
\end{table}

All simulations reported in this paper were performed with a two-dimensional ($D=2$) LB model of order $N = 3$
using a constant value of the fluid temperature $T$. 
For convenience, in Table \ref{roots} we provide the Cartesian projections
of the $16$ velocity vectors ${\bm{\xi}}_{k}$
used in this model, as well as their associated weights $w_{k}$
\cite{shan_jfm2006,dsfd2014}. More than a decade ago,
this $16$ velocity set was used also in entropic LB models 
\cite{ansumali_epl2003,bardow_epl2006,bardow_pre2008}.


\subsection{Force term}


The following expression of the acceleration ${\bm{g}}$ is used
in order to simulate the evolution of a van der Waals fluid where
the surface tension is controlled by the parameter $\kappa$
\cite{chen_annurev1998,succi_2001,sukop_2006,dsfd2014,pre2004,cicp2010,luo1998prl,
luo2000,coclite} :
\begin{equation}
{\bm{g}} \,=\,\frac{1}{\,\rho\,}\,{\bm{\nabla}}(p^{i}\,-\,p^{w}) \,+\,
\kappa {\bm{\nabla}} (\Delta \rho) 
\label{eder}
\end{equation}
where
$p^{i} = \rho T$ is the ideal gas pressure and 
$p^{w}$ is the van der Waals pressure given in Eq.~\eqref{evdw} below.
The equilibrium properties of the fluid can be described by the
Helmholtz free-energy functional \cite{rowl}
\begin{equation}
\Psi=\int d{\bf x} \Big [ \psi(\rho,T) + \frac{\kappa}{2}(\nabla \rho)^2
\Big ]
\label{free-en}
\end{equation}
where the bulk free-energy density is
\begin{equation}
\psi= \rho T \ln \Big (\frac{3 \rho}{3-\rho} \Big )-\frac{9}{8}\rho^2 .
\label{free-en2}
\end{equation}
The pressure tensor ${\bm \Pi}$ \cite{evans}
can be computed from Eq.~(\ref{free-en})
\begin{equation}
{\bm \Pi}=\Big [ p^w-\kappa \rho \Delta \rho 
-\frac{\kappa}{2} (\nabla \rho)^2 \Big ] {\bm 1}
+\kappa {\bm \nabla} \rho {\bm \nabla} \rho
\label{pr-tensor}
\end{equation}
Here ${\bm 1}$ is the unit tensor and
\begin{equation}
p^w=\rho \frac{\partial \psi}{\partial \rho}-\psi=
\frac{3\rho T}{3-\rho} - \frac{9}{8}\,\rho^{2}
\label{evdw}
\end{equation}
is the non-dimensionalized van der Waals equation of state
 with the critical point at $\rho_c=1$, $T_c=1$.
The acceleration ${\bf g}$ is then related to the pressure tensor
by the relationship
\begin{equation}
\rho {\bm g}={\bm \nabla} p^{i} - {\bm \nabla} \cdot {\bm \Pi} .
\end{equation}
In the presence of the force term $F_{k}$ given by Eq.~\eqref{herforce},
the conservation equations for mass and momentum, as derived
from (\ref{eBGK}) using the Chapman-Enskog procedure, are
\cite{klimontovich,pre2004,cicp2010,noitermico}
\begin{eqnarray}
\partial_{t}\rho + \nabla (\rho {\bm{u}}) & = & 0 \\
\partial_{t}(\rho {\bm{u}}) + \nabla (\rho {\bm{u}}{\bm{u}})
& = & - \nabla \cdot \left[ \bm{\Pi} - {\bm{S}}\right]
\rule{0mm}{7mm}
\end{eqnarray}
where the components of the viscous stress tensor ${\bm{S}}$ are
\begin{equation}
S_{\alpha\beta} = \rho T\tau \,\left[ \partial_{\alpha} u_{\beta} +
 \partial_{\beta} u_{\alpha} - (\nabla\cdot {\bm{u}}) \delta_{\alpha\beta}
\right] .
\label{estress}
\end{equation}
Unlike the LB models of order $N=2$, the term 
$(\nabla\cdot {\bm{u}}) \delta_{\alpha\beta}$
of the viscous stress tensor in Eq.~\eqref{estress}, is no longer
neglected in the present model and no spurious terms appear.

The use of large stencils in order to compute the space derivatives
of the pressure difference $(p^{i}\,-\,p^{w})$ and the local fluid density
$\rho$, which appear in Eq.~\eqref{eder},
is known to improve the isotropy of the phase interface, as well as
the accuracy of the values of the coexistence densities in the phase
diagram \cite{shan2008,leclaire,kart,mattila2014,siebert2014,dsfd2014}.
In this paper,  we used a {{25 point stencil}} to compute the values of
${\bm{\nabla}}(p^{i}\,-\,p^{w})$ and $\kappa {\bm{\nabla}} (\Delta \rho)$. The
procedure is documented in
Refs.\cite{dsfd2014,leclaire,kart,mattila2014,siebert2014} and
can be easily implemented on Graphics Processing Units
(GPUs) using the shared memory facility
\cite{farber,cook,professionalCUDA,cudaguide}.


\subsection{Corner transport upwind schemes}



\subsubsection{First order corner transport upwind}


The $16$ velocity vectors ${\bm{\xi}}_{k}$,  whose Cartesian projections
are shown in Table \ref{roots}, are {\emph{off-lattice}} vectors,
i.e., vectors that do not point from one node of the square
lattice to another one. For this reason, the
{\emph{collision - streaming}} scheme
\cite{chen_annurev1998,succi_2001,sukop_2006,aidun_annurev2010}
cannot be used in this case. Alternative schemes like the
interpolation supplemented LB schemes, the Runge-Kutta
time-marching schemes associated with various space-discretization
methods, or the elaborate characteristics-based off-lattice LB schemes
\cite{yuan2006,ansumali_epl2003,bardow_epl2006,bardow_pre2008,deville,philippi_pre2006,siebert_pre2008,surmas_eurJph2009,chika_pre2009,ansumali_pre2008,ansumali_pre2010,islb1997,islb2004,jcph2009,ubertini_ccph2008,guo_pre2003,lee_jcph2001,lee_jcph2003,hejranfar_pre2015}
are computationally expensive and difficult to stabilize, besides requiring 
specific treatment of the force and the
advection terms in the evolution equations (\ref{eBGK}). 

The {\emph{first order corner transport upwind}} (CTU1) scheme was
introduced more than two decades ago in the
mathematical literature related to hyperbolic equations
\cite{colella1990,leveque1996,leveque2002,trangenstein2009}.
Although this scheme is simple
enough and very convenient for solving the LB evolution
equations (\ref{eBGK}) on square or cubic lattices, regardless of the
orientation of the velocity vectors ${\bm{\xi}}_{k}$, its application
to LB models was not considered in the literature
until recently \cite{dsfd2014,australia2014}. Other
finite-volume schemes, mainly developed for
non-uniform meshes, were already used in the so-called
{\emph{volumetric lattice Boltzmann models}}
\cite{nannelli1992,hchen_pre1998,rzhang_pre2001,mauro_pre2010}.

The evolution of $f_{k} \equiv f_k({\mathbf{x}},t)$
is governed by Eqs.~\eqref{eBGK}, which
form a system of hyperbolic equations with non vanishing source terms.
A simple way to solve hyperbolic equations with source terms
is to split them into two steps, which can be treated explicitly \cite{leveque2002}. The first step refers to the advection
process, i.e., the left hand side of Eq.~\eqref{eBGK}, while the second
one refers to its right hand side, which includes the collision term as well
as the force term.
Let us consider the lattice cell centered in the node ${\mathbf{x}} =(x,y)$
of a $2D$ square lattice with $L_{x} \times L_{y}$ nodes.
For convenience, we introduce the notation
$f_{k,i,j}^{n} \equiv f_{k}(x = i \delta s, y = j \delta s, t=n\delta t)$, where
$\delta s$ is the lattice spacing, 
$0 \leq i < L_{x}$, $0 \leq j < L_{y}$, 
$\delta t$ is the time step and 
$n = 0,\, 1,\, 2,\, \ldots \infty$.
When using the CTU1 scheme to account
for the advection process, the Courant-Friedrichs-Levy
(CFL) condition \cite{trangenstein2009}
\begin{equation}
 {\mathrm{max}}_{k} \,
 \left\{\,|\xi_{k,x}|\delta t\, ,\, |\xi_{k,y}|\delta t\,\right\}\,
  \leq \,\delta s
  \end{equation}
ensures that
the new value $f_{k,i,j}^{n+1}$ receives
contributions from at most four neighboring nodes, according to
\cite{dsfd2014,trangenstein2009,australia2014}
\begin{eqnarray}
f_{k,i,j}^{n+1} \, = \, &
\frac{{\displaystyle{1}}}
{\,{\displaystyle{( \delta s)^{2}}}\,\rule{0mm}{4mm}} &
\left[ \rule{0mm}{4mm}\,
{{f_{k,i,j}^{n}\,
(\delta s - \vert\xi_{k,x}\vert \delta t)\,(\delta s - \vert\xi_{k,y}\vert \delta t)}}
\, + \, f_{k,i-\varsigma_{k,x},j}^{n} \,
\vert\xi_{k,x}\vert \, (\delta s - \vert\xi_{k,y}\vert \delta t)\, \delta t \,
 \right.  \nonumber \\
&  + & \left.
\,
f_{k,i,j-\varsigma_{k,y}}^{n}\, \vert\xi_{k,y}\vert \,
(\delta s - \vert\xi_{k,x}\vert \delta t)\, \delta t 
\, + \, f_{k,i-\varsigma_{k,x},j-\varsigma_{k,y}}^{n} \,
\vert\xi_{k,x}\vert \,  \vert\xi_{k,y}\vert \, (\delta t)^{2} \,
\rule{0mm}{4mm}\right]   \label{ctueq}
\rule{0mm}{8mm} 
\end{eqnarray}
In the equation above, the symbol $\varsigma_{k,\alpha}$, $1\leq k \leq K$,
 $\alpha\in \{x,y\}$, is defined as follows:
 \begin{equation}
\varsigma_{k,\alpha}  = \left\{ \begin{array}{rcl}
1  & , & \xi_{k,\alpha}  \geq 0 \\
- 1  & , & \xi_{k,\alpha}  < 0 . 
\end{array} \right.
\end{equation}
Note that $\xi_{k,\alpha}\,=\,\varsigma_{k,\alpha}\vert \xi_{k,\alpha}\vert$,
where $\vert \xi_{k,\alpha}\vert$ is the modulus of $\xi_{k,\alpha}$
(the sum rule over repeated indices is not considered for the symbol
$\varsigma_{k,\alpha}$). Figure 1 in Ref.~\cite{dsfd2014}, as well as
 Figure 2 in Ref.~\cite{australia2014},  illustrate the application of the
 CTU1 scheme \eqref{ctueq}
 when $\xi_{k,x} > 0$ and $\xi_{k,y} > 0$. In this case,
specific fractions of the neighboring distribution functions
 $f_{k,i-1,j}^{n}$, $f_{k,i-1,j-1}^{n}$ and $f_{k,i,j-1}^{n}$
are transported to the cell
$(i,j)$ across the sides of its lower left corner and contribute to
$f_{k,i,j}^{n+1}$, besides the remaining fraction of $f_{k,i,j}^{n}$.
  
Expanding
$f_{k,i,j}^{n+1}$,
$f_{k,i-\varsigma_{k,x},j}^{n}$,
$f_{k,i,j-\varsigma_{k,y}}^{n}$ and
$f_{k,i-\varsigma_{k,x},j-\varsigma_{k,y}}^{n}$
in Eq.~\eqref{ctueq} up to second order with respect to
$\delta s$ and $\delta t$, we get
\begin{eqnarray}
\partial_{t}f_{k} \,+\,\frac{1}{\,2\,}\,\delta t\,\partial_{t}^{2}f_{k}
\,+\,{\bm{\xi}}_{k}\cdot \nabla f_{k} & = & \nonumber \\
+\,\frac{1}{\,2\,}\,\delta s\, \vert\xi_{k,x}\vert   \,\partial_{x}^{2} f_{k} \,+\,
\frac{1}{\,2\,}\,\delta s\, \vert\xi_{k,y}\vert  \,\partial_{y}^{2} f_{k} \,+\,
\delta t\,\xi_{k,x} \xi_{k,y}\,\partial_{x}\partial_{y} f_{k} .
\label{eqevol2}
\end{eqnarray}
In order to get rid of the second order time derivative, we differentiate
Eq.~\eqref{eqevol2} with respect to time and retain only the terms up to
second order in $\delta s$ and $\delta t$:
\begin{equation}
\partial_{t}^{2} f_{k} \,=\,
\left[\, \xi_{k,x}^{2}\partial_{x}^{2}f_{k} \,+\,
\xi_{k,y}^{2} \partial_{y}^{2} f_{k}\,+\,
2 \xi_{k,x}\xi_{k,y} \partial_{x}\partial_{y}f_{k} \,\right] .
\end{equation}
Thus, the final form of the evolution equations solved using the CTU1
scheme is, up to second order in $\delta s$ and $\delta t$,
\begin{eqnarray}
 \partial_{t} f_{k} \,+\, \xi_{k,\gamma} \,\partial_{\gamma} f_{k} & = &
  -\,\frac{1}{\,\tau\,}\,\left[\, f_{k}\,-\,f^{eq}_{k}\,\right]\,+\, F_{k}
 \nonumber \\
& + & \frac{1}{\,2\,}\,\delta s\,\left[\,\vert\xi_{k,x}\vert
  \left(1-\vert\xi_{k,x}\vert\frac{\,\delta t\,}{\delta s}\right)
  \partial_{x}^{2} f_{k}
\,+\, \vert\xi_{k,y}\vert
 \left( 1-\vert\xi_{k,y}\vert\frac{\,\delta t\,}{\delta s}\right)
 \partial_{y}^{2} f_{k}  \,\right] .\rule{0mm}{9mm}
 \label{ctuevol3}
\end{eqnarray}
The last term in the square brackets of the equation above contributes
to the numerical viscosity \cite{jcph2003}. One can easily see that the
{\emph{collision-streaming}} scheme,
which is widely used in the two-dimensional D2Q9 LB model
\cite{chen_annurev1998,succi_2001,sukop_2006,aidun_annurev2010,shan_jfm2006},
is a particular case of the CTU1 scheme (\ref{ctuevol3}). The D2Q9 model
has nine ${\emph{on-lattice}}$ velocity vectors ${\bm{\xi}}_{k}$,
whose Cartesian projections
$\xi_{k,x}$ and $\xi_{k,y}$  take the values
$0$ or $\delta s / \delta t$.


\subsubsection{Second order corner transport upwind}


The {\emph{second order corner transport upwind}} (CTU2) scheme
improves the accuracy of the CTU1 scheme \eqref{ctueq}
by using flux limiters. Detailed description of this very elaborated scheme
can be found in Refs.\cite{leveque1996,leveque2002,trangenstein2009}.
A summary is given below.

Following Ref.\cite{leveque1996},  one defines the auxiliary variables
\begin{subequations}
\begin{eqnarray}
R_{k,i,j}^{x,n} & = & f_{k,i,j}^{n} - f_{k,i-1,j}^{n} , \\
R_{k,i,j}^{y,n} & = & f_{k,i,j}^{n} - f_{k,i,j-1}^{n} ,
\rule{0mm}{7mm} 
\end{eqnarray}
\end{subequations}
\begin{subequations}
\begin{eqnarray}
S_{k,i,j}^{x,n} & = & \frac{1}{\,2\,} \vert \xi_{k,x} \vert
\left( 1 - \frac{\delta t}{\,\delta s\,} \vert \xi_{k,x} \vert \right)
R_{k,i,j}^{x,n}  \Psi
\left( \frac{\,R_{k,i-\varsigma_{k,x},j}^{x,n}\,}{R_{k,i,j}^{x,n}} \right) ,
\\
S_{k,i,j}^{y,n} & = & \frac{1}{\,2\,} \vert \xi_{k,y} \vert
\left( 1 - \frac{\delta t}{\,\delta s\,} \vert \xi_{k,y} \vert \right)
R_{k,i,j}^{y,n}  \Psi
\left( \frac{\,R_{k,i,j-\varsigma_{k,y}}^{x,n}\,}{R_{k,i,j}^{x,n}} \right) , 
\rule{0mm}{10mm}
\end{eqnarray}
\end{subequations}
where $\Psi (\theta)$ is a flux limiter. In this paper
we will use the monitorized centered limiter (MC) \cite{leveque1996,leveque2002,trangenstein2009,cejp2004}
\begin{equation}
\Psi (\theta) =
{\mathrm{max}}\left\{\,0,\, {\mathrm{min}}[\, (1+\theta)/2,\, 2,\, 2\theta\,]\,\right\} .
\end{equation}
The fluxes ${\mathcal{F}}_{i+1/2,j}$ and ${\mathcal{G}}_{i,j+1/2}$,
which exit the cell $(i,j)$ in the $x$ and $y$ directions, respectively, are
defined by
\begin{subequations}
\begin{eqnarray}
{\mathcal{F}}_{k, \, i+\frac{1}{2}  ,\, j}^{n} & = &
f_{k , \, i+\frac{1}{2}(1-\varsigma_{k,x})  , \, j} \, \xi_{k,\,x} 
\, + \, S_{k, \, i+1 , \,j}^{x,n} 
\nonumber \\
& - & \frac{1}{2}\,\frac{\delta t}{\,\delta s\,} \, \xi_{k, \, x}\, \xi_{k, \, y} \,
R_{k, \, i+\frac{1}{2}(1-\varsigma_{k, \,y}),
 \,j+\frac{1}{2}(1-\varsigma_{k,\,y})}^{y, \,n}
\rule{0mm}{8mm} \nonumber \\
 & + & \frac{\delta t}{\,\delta s\,}\,\xi_{k,\,x} \,
\left[\,S_{k, \, i+\frac{1}{2}(1-\varsigma_{k,\, x}), \,j}^{y,\,n} -
 S_{k, \, i+\frac{1}{2} (1-\varsigma_{k,\,x}) , \, j+1}^{y,\,n} 
\, \right] , \rule{0mm}{8mm}
\\
{\mathcal{G}}_{k, \, i  ,\, j+\frac{1}{2}}^{n} & = &
f_{k , \, i  , \, j+\frac{1}{2}(1-\varsigma_{k,\,y})} \, \xi_{k,\,y} 
\, + \, S_{k, \, i , \,j+1}^{y,\,n} 
\rule{0mm}{8mm}\nonumber \\
& - & \frac{1}{2}\,\frac{\delta t}{\,\delta s\,} \, \xi_{k, \, x}\, \xi_{k, \, y} \,
R_{k, \, i+\frac{1}{2}(1-\varsigma_{k, \,x}),
 \,j+\frac{1}{2}(1-\varsigma_{k,y})}^{x, \,n}
\rule{0mm}{8mm} \nonumber \\
 & + & \frac{\delta t}{\,\delta s\,}\,\xi_{k,\,y} \,
\left[\,S_{k, \, i, \,j+\frac{1}{2}(1-\varsigma_{k,\, y})}^{x,\,n} -
 S_{k, \, i+1  , \, j+\frac{1}{2} (1-\varsigma_{k,\,y})}^{x,\,n} 
\, \right] . \rule{0mm}{8mm}
\end{eqnarray}
\end{subequations}
The incoming numerical fluxes 
${\mathcal{F}}_{i-1/2,j} \equiv {\mathcal{F}}_{(i-1)+1/2,\,j}$
and ${\mathcal{G}}_{i,\,j-1/2} \equiv {\mathcal{G}}_{i,\,(j-1)+1/2}$
are defined in a similar manner.
According to the CTU2 scheme, the distribution function $f_{k,\,i,\,j}^{n}$
is updated as follows \cite{leveque1996,leveque2002,trangenstein2009} :
\begin{eqnarray}
f_{k,\,i,\,j}^{n+1} & = & f_{k,\,i,\,j}^{n}\,-\, \frac{\delta t}{\,\delta s\,}\,
[\,{\mathcal{F}}_{i+1/2,\,j} \,-\, {\mathcal{F}}_{i-1/2,\,j} \,+\,
{\mathcal{G}}_{i,\,j+1/2} \,-\, {\mathcal{G}}_{i,\,j-1/2} \,] \\
& & -\,\frac{\,\delta t\,}{\,\tau\,}\,\left[\, 
f_{k,\,i,\,j}\,-\,f^{eq}_{k,i,j}\,\right]\,+\,
\delta t F_{k} \rule{0mm}{8mm} \nonumber
\end{eqnarray}
where $f^{eq}_{k,\,i,\,j}$ and $F_{k,\,i,\,j}$ are calculated according to
Eqs.~(\ref{herfeq}) and (\ref{herforce}), respectively.
In this case an analytical expression for the numerical
viscosity cannot be derived but it is supposed to be at the second
order in the lattice spacing $\delta s$.


\subsection{Numerical viscosity, anisotropy and regularization}
\label{nvsection}


In order to investigate possible anisotropy due to numerical effects,
in this subsection we analyze the evolution of
shear waves of wavelength $\lambda = 2$  in an ideal gas with density
$\rho=1$ at temperature 
$T=1$ by setting $F_{k}=0$, $1 \le k \le K$,
in the evolution equation \eqref{eBGK}.
Computer simulations were performed using both the CTU1 and the CTU2
numerical schemes on a two-dimensional square lattice with
$L_{x} \times L_{y}$ nodes along the Cartesian axes,
where periodic boundary conditions apply.
For each numerical scheme, we conducted two series of simulations with
the time step $\delta t = 10^{-4}$. In the first series, the wave vector
${\bm{k}}$, $\vert {\bm{k}} \vert = 2\pi/\lambda$, was aligned 
along the horizontal axis of the square lattice and its Cartesian components
were $(2\pi / \lambda, 0)$. This series will be denoted as
the axial (A) one. In the second series, denoted as the diagonal (D) one,
the wave vector ${\mathbf{k}}$ was aligned along the diagonal direction
of the square lattice after a counterclockwise rotation by an angle $\pi/4$, 
hence its Cartesian components were
$(\pi\sqrt{2}/\lambda,\pi\sqrt{2}/\lambda)$.

To account for the numerical effects induced by the CTU1 and CTU2 schemes,
two values of the lattice spacing $\delta s$ were used in
each series, namely $1/128$ and $1/256$. When
conducting the first series of simulations with these values of $\delta s$,
the  wavelength $\lambda=2$ of the shear waves was easily secured on
lattices with $256\times 256$ and $512\times 512$ nodes, respectively.
To match the periodic boundary conditions for $\lambda = 2$
using the same values of $\delta s$ when simulating the diagonal waves,
we conducted the simulations on square lattices with
$362\times 362$ and $724\times 724$ nodes, respectively, as suggested
in \cite{zhang2006pre}.

Let ${\bm{u}}({\bm{x}},t)$ be the fluid velocity vector in the node
${\bm{x}} = (i\delta s, j\delta s)$ of the lattice at time $t$. 
The components of
the vector ${\bm{u}}({\bm{x}},t)$, which are parallel or perpendicular to the
wave vector $\bm{k}$, are denoted $u_{\parallel}({\bm{x}},t)$ and
$u_{\perp}({\bm{x}},t)$, respectively.
In both the series of simulations, the shear waves were
initialized according to:
\begin{subequations}
\begin{eqnarray}
u_{\parallel}({\bm{x}},0) & = & 0 \\
u_{\perp}({\bm{x}},0) & = & U \cos({\bm{k}} \cdot {\bm{x}}) 
\end{eqnarray}
\end{subequations}
with $U = 0.01$. When the fluid is not too far from the equilibrium (i.e.,
when the relaxation time is small enough), the fluid evolves according
to the Navier-Stokes equations. For shear waves, we have
$u_{\parallel}({\mathbf{x}},t) = 0$ and there is no
spatial variation of the velocity vector along the direction perpendicular to
the wave vector. Under these circumstances, and assuming that the fluid is isothermal and incompressible, the shear wave equation reads
\begin{equation}
\partial_{t} u_{\perp}({\bm{x}},t)\,-\,\nu_{app}\,
\partial_{\parallel}^{2}u_{\perp}({\bm{x}},t) = 0
\end{equation}
where $\nu_{app}$ is the {\emph{apparent value}} of the kinematic
viscosity \cite{jcph2003} and $\partial_{\parallel}^{2}$ denotes the
second order space derivative along the direction of the wave vector.
As described in \cite{jcph2003}, the value of the apparent viscosity
can be determined at time $t$ according to
\begin{equation}
\nu_{app} \,=\ \frac{1}{\,k^{2}t\,}\,
\log\,\frac{\,u_{\perp}(0,0)\,}{\,u_{\perp}(0,t)\,}
\label{avisco}
\end{equation} 
where $k=\vert{\bm{k}}\vert$.

\begin{figure}[H]
\begin{center}
\includegraphics*[width=.9\textwidth]{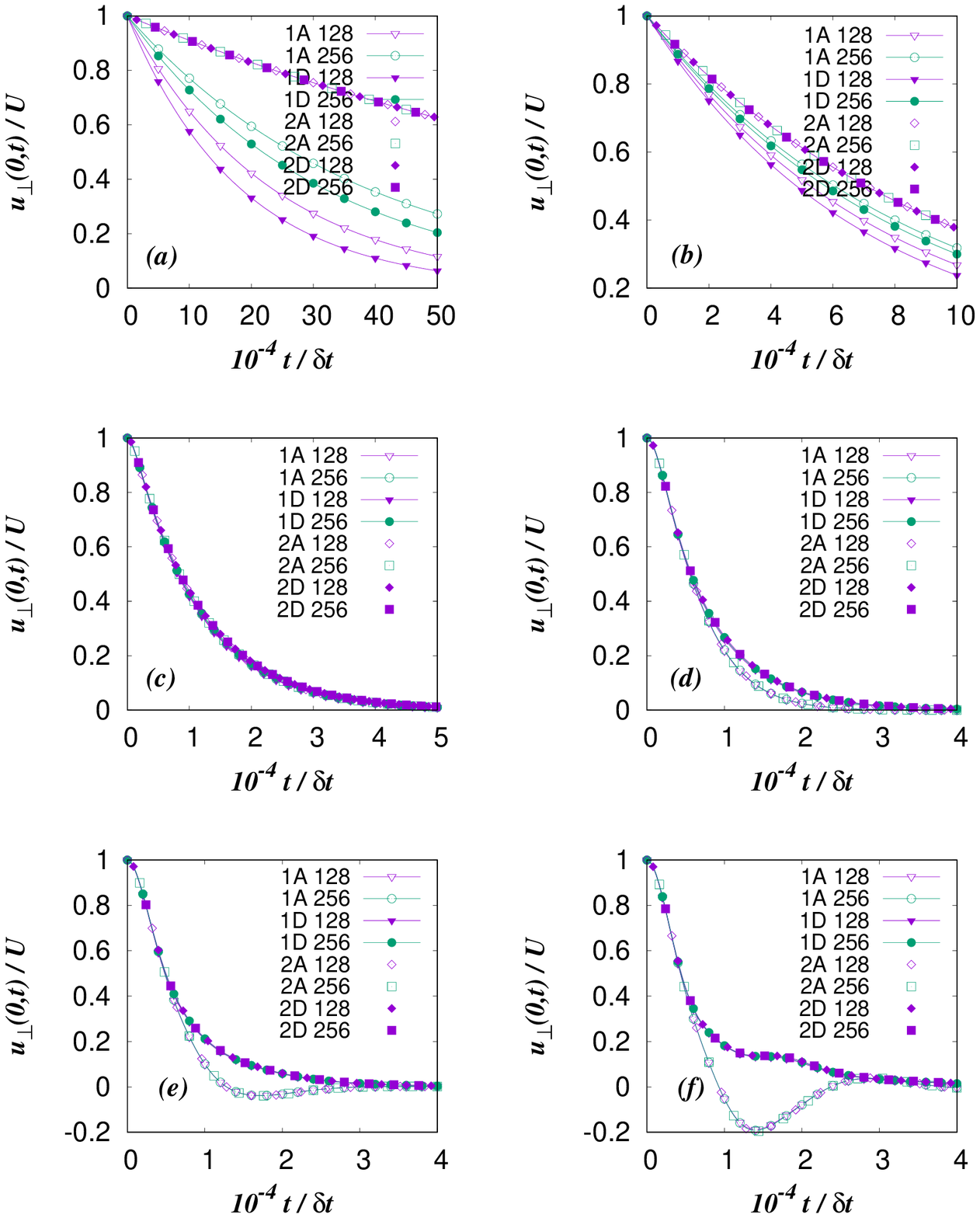} 
\end{center}
\caption{Evolution of the normalized peak velocity $u_{\perp}(0,t)/U$
of decaying shear waves without regularization, for various values of
the relaxation time $\tau=$ 0.001 (a), 0.010 (b), 0.100 (c), 0.200 (d), 
0.300 (e), 0.500 (f). The results obtained
using the CTU$n$ scheme, $n\in \{ 1,2\}$, and the lattice spacing
$\delta s = 1/S$, $S \in\{128,256\}$, for axial (A) and diagonal waves (D),
are marked with $n$A~$S$ and $n$D~$S$, respectively. 
}
\label{wfiguregnew}
\end{figure}

Figure \ref{wfiguregnew} shows the evolution
of the normalized peak velocity $u_{\perp}(0,t) / U$ for
six values of the relaxation time
$\tau$.
When
using the CTU1 scheme and small values of the relaxation time
($\tau = 0.001, 0.01$), the evolution of the shear waves
in the two directions (axial and diagonal) differs significantly for both
values of the lattice spacing
$\delta s$ considered in our simulations.
This is due to the anisotropy of the numerical effects, which plague the 
solutions of hyperbolic partial differential equations in multi-dimensional
spaces \cite{sescu2014,sescu2015}. The numerically induced anisotropy 
reduces significantly when using higher order schemes, as seen in
Fig.~\ref{wfiguregnew}, where the evolution of the shear waves
orientated along both the axial and the diagonal direction is quite identical
when using the CTU2 scheme with $\tau = 0.001, 0.01$.
Although both the CTU1 and the CTU2 simulations give close results
for $\tau = 0.1$, regardless of the orientation of the shear waves
or of the value of the lattice spacing $\delta s$, Fig.~\ref{wfiguregnew} shows that the evolution of the axial and the diagonal shear waves differ
again when $\tau$ is further increased. More precisely, when $\tau > 0.1$,
the evolution of the shear waves becomes more and more anisotropic and,
apparently, it no longer depends
either on the order of the CTU scheme
used to conduct the simulation or on the lattice spacing $\delta s$.
This kind of anisotropy, which manifests for higher values of $\tau$,
regardless of the numerical scheme used to evolve the distribution functions
$f_{k}$, can be reduced by using a regularization procedure, as will be 
discussed further in this subsection.

In order to understand all the features mentioned above, we
refer to Ref. \cite{jcph2003}, where it is assumed that the apparent
value $\nu_{app}$ of the kinematic viscosity of a fluid, observed
during simulations conducted with finite-difference LB models, is always the sum of two terms, the physical (theoretical) value of the viscosity
$\nu_{phys}$ and the numerical viscosity $\nu_{num}$
\begin{equation}
\nu_{app} \,=\,\nu_{phys} \,+\,\nu_{num} .
\label{evisco}
\end{equation}
When the fluid satisfies 
the Navier-Stokes equations, it can be shown that the application of the
Chapman - Enskog method \cite{jcph2003} gives 
\begin{equation}
\nu_{phys} = \rho\tau T
\label{phvisco}
\end{equation}
which is a constant quantity in the case of
our shear wave simulations. 
Table \ref{apparentvisco} shows the values of the apparent viscosity
$\nu_{app}^{axial}$ and $\nu_{app}^{diagonal}$,
as determined at $t=20$ using Eq.~\eqref{avisco} when using the CTU1
and the CTU2 schemes to simulate the shear wave decay with
$\tau \in \{0.001, 0.01\,\}$. For convenience, in 
Table \ref{numericalvisco} we show also the corresponding values of the
numerical viscosity, derived from Table \ref{apparentvisco}
according to Eqs.~\eqref{evisco} and \eqref{phvisco}, in the case
of the CTU1 scheme.  

Inspection of the
results in Table \ref{numericalvisco} reveals that the numerical
viscosity of the CTU1 scheme is practically independent of
the relaxation time $\tau$ and depends only on the orientation of the
shear waves, as well as on the lattice spacing $\delta s$. For each 
orientation (axial or diagonal) of the shear waves, it is easy to observe that
\begin{equation}
\nu_{num}^{orientation}({\mathrm{CTU1}},\tau,\delta s = 1/128)\, / \,
\nu_{num}^{orientation}({\mathrm{CTU1}},\tau,\delta s = 1/256) \simeq 2
\end{equation} 
This agrees with Eq.~\eqref{ctuevol3}, where the spurious (last) term
depends linearly on $\delta s$. Moreover, for both values of
$\delta s$ in Table \ref{numericalvisco}, one can see that
\begin{equation}
\nu_{num}^{diagonal}({\mathrm{CTU1}},\tau,\delta s)\, / \,
\nu_{num}^{axial}({\mathrm{CTU1}},\tau,\delta s) \simeq \sqrt{2}
\end{equation}
which is not a surprise since the distance between the lattice nodes
along the diagonal direction of the lattice is $\delta s \sqrt{2}$.
As the value of $\tau$ increases, the relative contribution of the numerical
viscosity $\nu_{num}$ to the apparent viscosity, Eq.~\eqref{evisco},
becomes smaller. This explains why the evolution of the axial and the
diagonal shear waves becomes quite identical, as seen in Fig.~\ref{wfiguregnew} when using the CTU1 scheme with $\tau = 0.1$.

The numerical effects introduced by the CTU2 scheme are much smaller
than in the case of the CTU1 scheme. For this reason, the evolution of
shear waves, as seen for $\tau \le 0.1$ in the CTU2 simulations reported
in Fig.~\ref{wfiguregnew}, is quite independent on their orientation,
as well as on the value of $\delta s$. Moreover, in Table \ref{apparentvisco} 
one can see that the CTU2 values of the apparent viscosity, reported
for $\tau = 0.001$ and  $\tau = 0.01$, are close enough to the corresponding
physical values given by Eq.~\eqref{phvisco}. 

\begin{table}
\caption{Apparent values of the viscosity of the shear waves
orientated along the axial or the diagonal direction of a square lattice,
calculated at $t=20$
using the first and the second order corner transport schemes,
for small values of the relaxation time $\tau$ and
two values of the lattice spacing $\delta s$.}
\label{apparentvisco}
\begin{center}
\begin{tabular}{cccc|cccc|ccccc}
\hline\hline 
& & & & \multicolumn{3}{c} {$\nu_{app}^{axial}$} & &
 \multicolumn{3}{c} {$\nu_{app}^{diagonal}$} \\
 \hline\hline
\quad  $\tau$ & &  $\delta s$  & \qquad & 
\quad CTU1  & & CTU2 & & \quad CTU1 & & CTU2 \\ \hline\hline
\quad 0.001 & \quad  & 1/128 & \qquad &
\quad 4.3685e-03 & &  9.5054e-04 \quad & &
\quad 5.5900e-03 & &  9.5024e-04 \qquad \\
& & 1/256 & &
\quad  2.6343e-03 & & 9.5002e-04 \quad & &
\quad 3.2201e-03 & &  9.4990e-04 \qquad \\ \hline
\quad 0.010 & & 1/128 & &
\quad  1.3344e-02 & & 9.9367e-03 \quad & &
\quad  1.4572e-02 && 9.9380e-03 \qquad\\
 & & 1/256 & &
\quad  1.1615e-02 & & 9.9356e-03 \quad & &
\quad 1.2205e-02 & & 9.9373e-03 \qquad \\ \hline\hline
\end{tabular}
\end{center}
\end{table}

\begin{table}
\caption{Numerical viscosities observed
during the simulation of shear waves with the CTU1 scheme
at small values of the relaxation time $\tau$, as calculated from
Table \ref{apparentvisco} by subtracting the corresponding values
of  $\nu_{phys} = \rho\tau T$.}
\label{numericalvisco}
\begin{center}
\begin{tabular}{cccc|cccc|ccccc}
\hline\hline 
\quad  $\tau$ & &  $\delta s$  & \qquad & 
\quad  $\nu_{num}^{axial}({\mathrm{CTU1}},\tau,\delta s)$ & & & & 
\quad $\nu_{num}^{diagonal}({\mathrm{CTU1}},\tau,\delta s)$ & &  \\
 \hline\hline
\quad 0.001 & \quad  & 1/128 & \qquad &
\quad 3.3685e-03 & &   & &
\quad 4.5900e-03 & &   \\
& & 1/256 & &
\quad  1.6343e-03 & &   & &
\quad 2.2201e-03 & &   \\ \hline
\quad 0.010 & & 1/128 & &
\quad  0.3344e-02 & &  & &
\quad  0.4572e-02 && \\
 & & 1/256 & &
\quad  0.1615e-02 & &  & &
\quad 0.2205e-02 & &  \\ \hline\hline
\end{tabular}
\end{center}
\end{table}

For $\tau > 0.1$, the plots in Fig.~\ref{wfiguregnew} show that
the evolution of the shear waves becomes more and more anisotropic
and does not depend either on the numerical scheme or on the lattice spacing
$\delta s$. This kind of anisotropy, which develops when the fluid system
lies further and further from the equilibrium state (i.e., when the relaxation
time $\tau$ becomes large enough) is present also in the collision-streaming 
LB models 
\cite{zhang2006pre,latt2006mcs,colosqui2009pof,colosqui2010pre,montessori2014pre,mattila_pof2017} and originates from the non-equilibrium
part of the distribution function, which overpasses the space
of the tensor Hermite polynomials up to order $N$, used in the model.

Let us assume that at time $t=0$, the functions $f_{k}$, which evolve
according to Eq.~\eqref{eBGK},  are
expressed as an expansion up to the order $N=3$
\begin{equation}
f_{k} \equiv f_{k}({\bm{x}},t) \, = \,
 w_{k} \, \sum_{\ell=0}^{N} \,
\frac{1}{\ell!}  \, {\bm{a}}^{(\ell)}_{\alpha_{1} \ldots \alpha_{\ell}} 
({\bm{x}},t)\,
{\bm{\mathcal{H}}}^{(\ell)}_{\alpha_{1} \ldots \alpha_{\ell}}({\bm{\xi}}_{k}) ,
\label{exfh2} 
\end{equation}
with respect to the tensor Hermite polynomials
${\bm{\mathcal{H}}}^{(\ell)}_{\alpha_{1} \ldots \alpha_{\ell}}({\bm{\xi}}_{k})$,
in a similar way as the expansion \eqref{exfeqh2} of $f^{eq}$.
Since the functions $f_{k}$ are subjected to the transport operator
${\bm{\xi}}_{k} \cdot{\bm{\nabla}}$ in the evolution equation \eqref{eBGK},
the application of the recurrence relation \cite{shan_jfm2006}
\begin{equation}
\xi_{\alpha}
{\bm{\mathcal{H}}}^{(\ell)}_{\alpha_{1} \ldots \alpha_{\ell}}({\bm{\xi}})
 \,=\,
{\bm{\mathcal{H}}}^{(\ell+1)}_{\alpha\alpha_{1} \ldots \alpha_{\ell}}({\bm{\xi}})  \,+\,
\sum_{k=1}^{\ell}\,\delta_{\alpha\alpha_{k}}\,
{\bm{\mathcal{H}}}^{(\ell-1)}_{\alpha_{1} \ldots \alpha_{k-1}\alpha_{k+1}\ldots\alpha_{\ell}} ({\bm{\xi}})
\label{e3}
\end{equation}
reveals that after the first time step the series expansion \eqref{exfh2}
of $f_{k}$ acquires a supplementary
term of order $N+1$. Subsequent time steps performed
during the computer simulation further increase the
order of the tensor Hermite polynomials in the expansion of $f_{k}$ and,
thus,  $f_{k}$ will lie outside the space where $f_{k}^{eq}$ are defined, that is,
the space generated by  the tensor Hermite polynomials up to a
certain order $N$ (e.g., $N=3$ as in this paper or $N=2$ as in the
D2Q9 LB model widely used in the literature). This behavior originates
from the recurrence property \eqref{e3}
of Hermite polynomials and is specific to any LB models based on the
Gauss-Hermite quadrature, including the one used in this paper.
However, when Cartesian projections of all the velocity vectors
${\bm{\xi}}_{k}$, $k=1,\,2,\,\ldots\,K$, used in the LB model
are roots of the Hermite polynomial $H^{Q}(\xi)$
of order $Q=N+1$, the tensor Hermite polynomials of order $N+1$
in Eq.~\eqref{e3} vanish
when all indices $\alpha,\,\alpha_{1},\,\ldots \alpha_{\ell=N}$ are
equal. This feature of the LB model
used in this paper, which does not allow the order of the
series expansion of $f_{k}$ to increase indefinitely
during the advection process \cite{ijmpf},
is further discussed in the Appendix.

It is known that the terms in the expansion \eqref{exfh2}
of  the distribution functions $f_{k}$, which
contain Hermite tensors of order higher than the order $N$ used in the
expansion of the equilibrium distribution functions $f_{k}^{eq}$, are
at the origin of numerous issues (numerical instabilities, anisotropy, low
accuracy, etc.) which manifest at higher values of the relaxation
time $\tau$
\cite{niu2007pre,suga2010pre,suga2013fdr,zhang2006pre,latt2006mcs,colosqui2009pof,colosqui2010pre,montessori2014pre,mattila_pof2017}.
To reduce these problems, one can use a regularization procedure
\cite{niu2007pre,suga2010pre,suga2013fdr,zhang2006pre,latt2006mcs,colosqui2009pof,colosqui2010pre,montessori2014pre,mattila_pof2017}.
Following this recipe, the non-equilibrium part
$f^{neq} = f_{k} - f_{k}^{eq}$ of the functions $f_{k}$, which enters the
BGK collision term in the evolution equation \eqref{eBGK}, is
replaced at each time step by \cite{zhang2006pre}
\begin{equation}
\hat{f}_{k}^{neq} \,=\, w_{k} \,\left[\,\frac{1}{\,2!\,} \,
{\bm{\mathcal{H}}}^{(2)}_{\alpha \beta}({\bm{\xi}}_{k})
\sum_{k'=1}^{K} f_{k'}^{neq}
{\bm{\xi}}_{k',\alpha}{\bm{\xi}}_{k',\beta}
\,+\,\frac{1}{\,3!\,} \,
{\bm{\mathcal{H}}}^{(3)}_{\alpha\beta \gamma}({\bm{\xi}}_{k})
\sum_{k'=1}^{K} f_{k'}^{neq}
{\bm{\xi}}_{k',\alpha}{\bm{\xi}}_{k',\beta}{\bm{\xi}}_{k',\gamma}\,\right]
\label{ereg}
\end{equation}
Application of the regularization procedure at every time step eliminates
the terms of order higher than $N=3$ in the Hermite expansion of
the distribution functions $f_{k}$, $k = 1,\,2,\,\ldots K$, hence both
$f_{k}$ and $f_{k}^{eq}$ remain in the space generated by the tensor
Hermite polynomials of order at most $N=3$.

\begin{figure}[H]
\begin{center}
\includegraphics*[width=.9\textwidth]{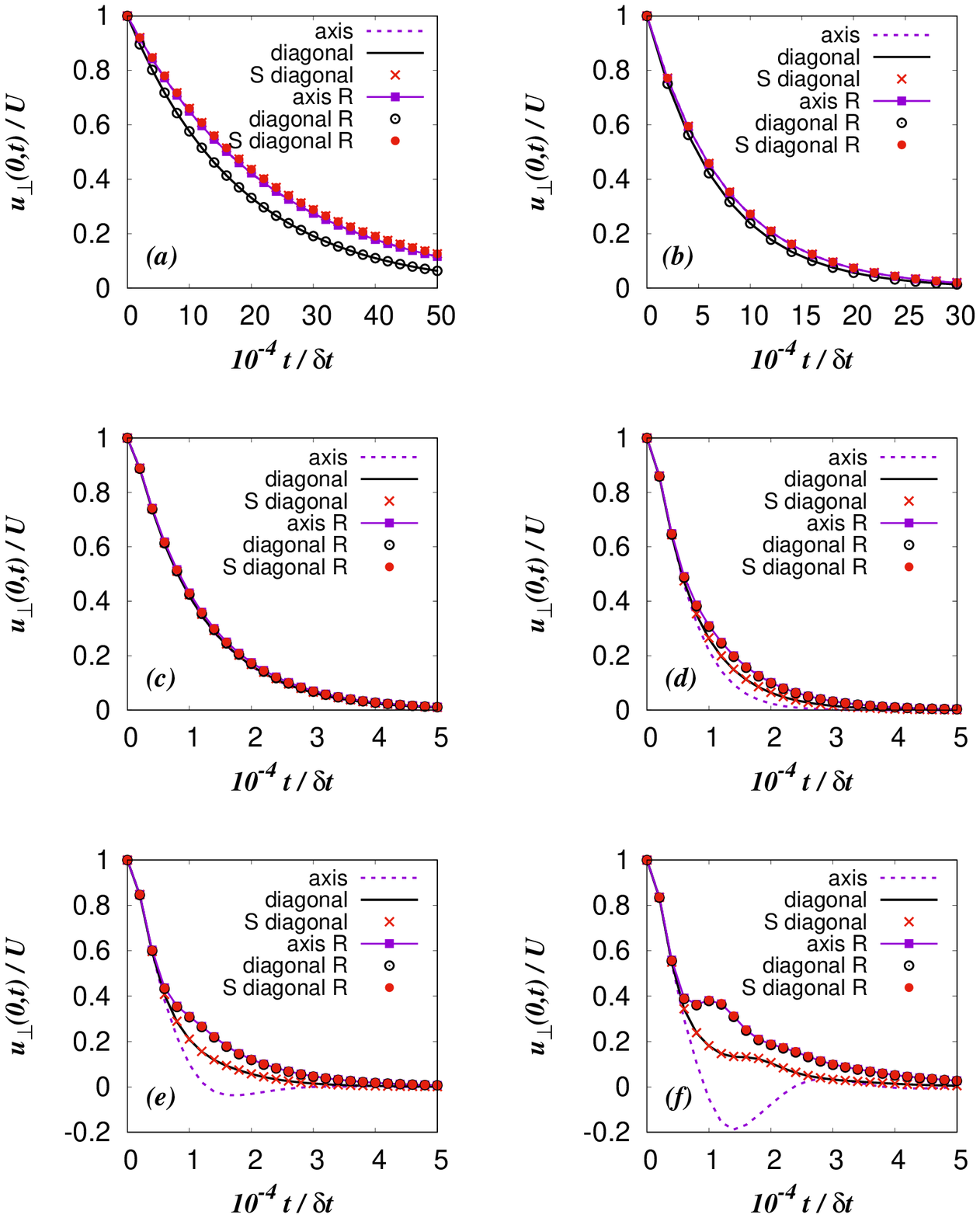} 
\end{center}
\caption{Evolution of the normalized peak velocity $u_{\perp}(0,t)/U$
of decaying shear waves obtained using the CTU1 numerical scheme
with and without regularization, for various values of the relaxation time
$\tau=$ 0.001 (a), 0.010 (b), 0.100 (c), 0.200 (d), 0.300 (e), 0.500 (f).
The results which carry the symbol S in the plot keys were obtained on
lattices with the spacing $\delta s=1/(128\sqrt{2})$, while the remaining results
were obtained on lattices with $\delta s = 1/128$. The results obtained using the regularization procedure are marked with the symbol R.}
\label{newfigureg}
\end{figure}

In Fig.~\ref{newfigureg}, we compare the evolution of the normalized
peak velocity $u_{\perp}(0,t)/U$ of shear waves of wavelength
$\lambda = 2$. For each value of the relaxation time $\tau$,
the plots in this figure show the decay of the normalized peak velocity
in three cases. 
In the first case, the wave vector $\mathbf{k}$ of the shear waves is
oriented along the horizontal axis of a square lattice lattice 
with spacing $\delta s = 1/128$. In the second and third cases,
the wave vector $\mathbf{k}$ is oriented along the diagonal of 
the square lattices with spacings  $\delta s = 1/128$ and
$\delta s = 1/(128\sqrt{2})$, respectively. The results obtained on the
lattice with the smaller spacing ($\delta s = 1/(128\sqrt{2})$) carry
the symbol S in the corresponding plot keys.
In all cases, the simulations were conducted using the CTU1 scheme with
or without  application of the regularization procedure, Eq.~\eqref{ereg} above.
The results obtained using the regularization procedure carry
the symbol R in the plot keys.

Inspection of the plots in Fig.~\ref{newfigureg} reveals that the application
of the regularization procedure does not change the evolution of shear
waves for $\tau \leq 0.1$, i.e., when the fluid is not far from the equilibrium.
Moreover, for $\tau < 0.1$ and $\delta s = 1/128$ one can see that
the axial and the 
diagonal shear waves evolve differently because of the anisotropy of the
spurious viscosity, as discussed previously.  Furthermore, for these
small values of $\tau$, the evolution of the diagonal waves on the
square lattice with $\delta s = 1/(128\sqrt{2})$ (the results marked with S
in the plot keys) agrees to the evolution of the axial waves
on the lattice with $\delta s = 1/128$, as expected since the numerical
viscosities are quite identical in these cases. For $\tau=0.1$, the evolution of
the shear waves is quite identical, regardless of their orientation or the
value of $\delta s$. As discussed previously, this happens because the
relative contribution of the numerical viscosity to the apparent value of the 
viscosity becomes negligible when $\tau$ is large enough.
When no regularization procedure is applied, the simulation results
for $\tau > 0.1$ become anisotropic again. Furthermore, one can see
that the evolution of the diagonal shear waves is identical, despite of
the different values of the lattice spacing $\delta s$. The application of the 
regularization procedure during the simulations fully
restores the isotropy, as
already known in the literature \cite{zhang2006pre,latt2006mcs,colosqui2009pof,colosqui2010pre,montessori2014pre,mattila_pof2017}.

\begin{figure}[H]
\begin{center}
\includegraphics*[width=.9\textwidth]{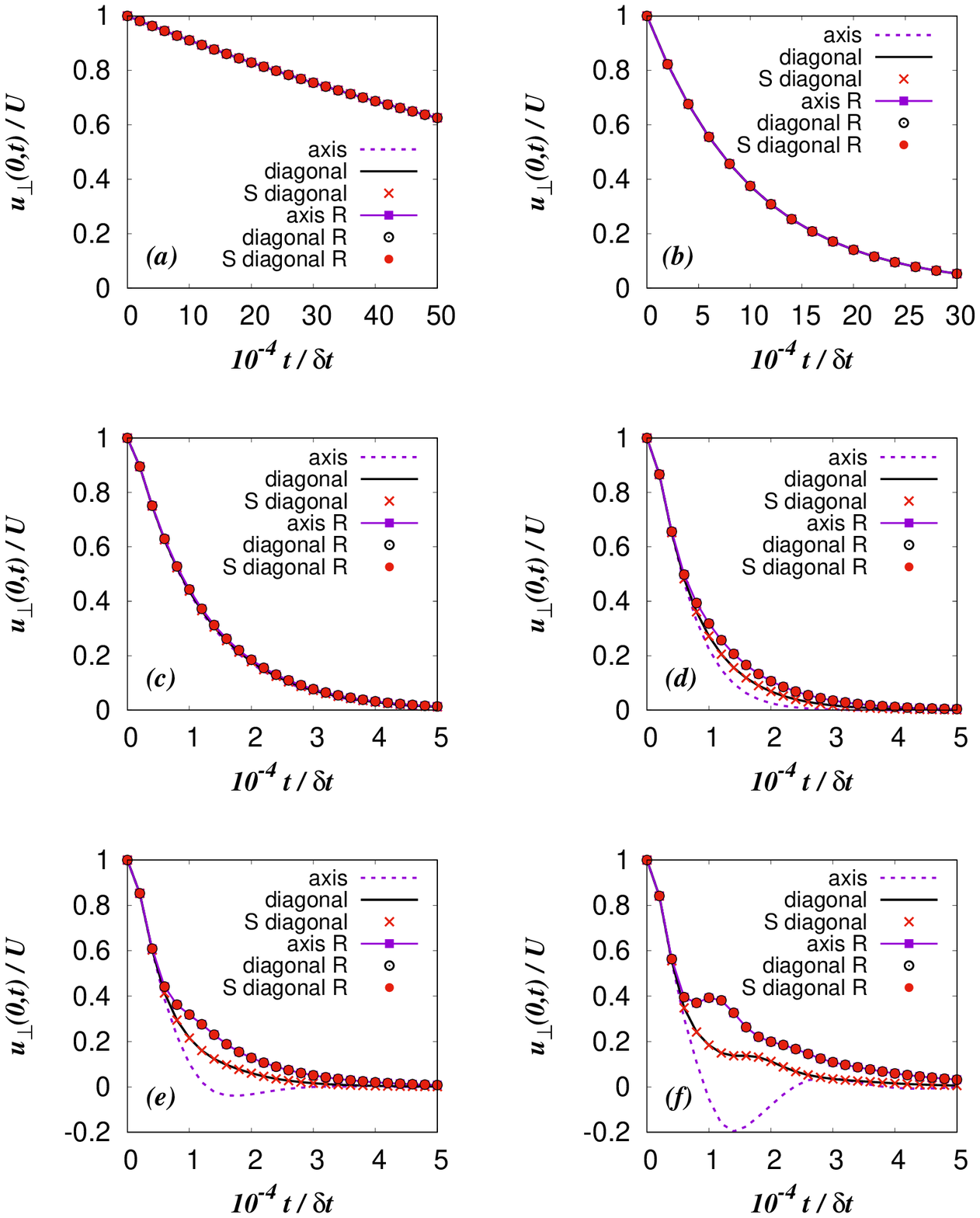} 
\end{center}
\caption{Evolution of the normalized peak velocity $u_{\perp}(0,t)/U$
of decaying shear waves obtained using the CTU2 numerical scheme
with and without regularization, for various values of the relaxation time
$\tau=$ 0.001 (a), 0.010 (b), 0.100 (c), 0.200 (d), 0.300 (e), 0.500 (f).
The results which carry the symbol S in the plot keys were obtained on
lattices with the spacing $\delta s=1/(256\sqrt{2})$, while the remaining results
were obtained on lattices with $\delta s = 1/256$. The results obtained using the regularization procedure are marked with the symbol R.}
\label{newfiguregg}
\end{figure}

We checked the regularization
also for the CTU2 scheme. 
The results shown in Fig.~\ref{newfiguregg}) 
confirm again that the application of the regularization procedure
cures the anisotropy which appears 
at large values of the relaxation time ($\tau > 0.1$).

Since the LB model introduced in this paper is used to investigate the
behavior of a cavitation bubble, which obeys the Navier-Stokes equations
for an isothermal fluid governed by the van der Waals equation of state,
Eq.~\eqref{evdw}, the values of the relaxation time $\tau$ to be
considered further during the
simulations need to be small enough ($\tau \le 0.01$) in order to ensure 
the correct recovery of these equations \cite{ambrus_pre2012,piaud_ijmpc2014,sspre2005,pre2014, sone,karniadakis}.
For this reason, we did not use the regularization
procedure during the simulations reported in Section \ref{cresults}
since it is not necessary, as just seen.


\subsection{Liquid - vapor phase diagram}
\label{nvsection2}

\begin{figure}[H]
\begin{center}
\includegraphics*[width=.70\textwidth]{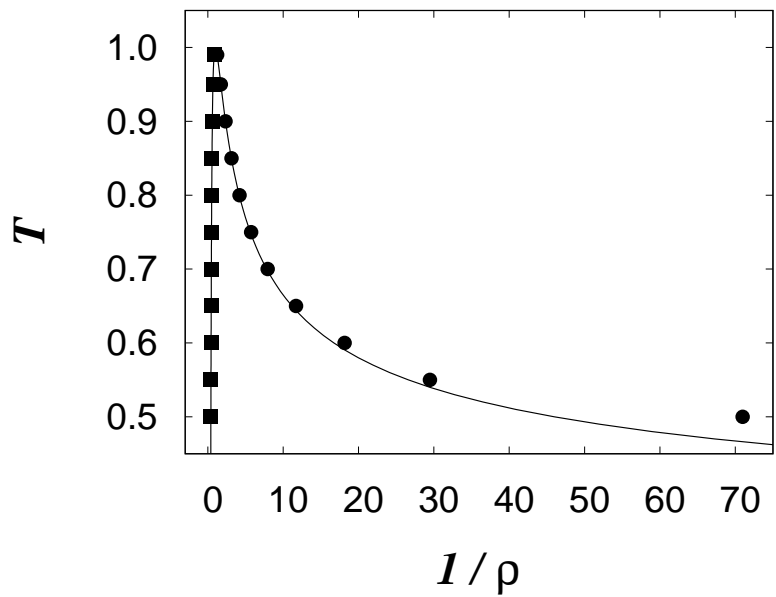} 
\end{center}
\caption{Liquid- vapor phase diagram: Symbols refer to LB results on the
liquid branch ($\blacksquare$) and on the vapor one ($\bullet$). The full lines
correspond to the results of the Maxwell construction.}
\label{fphase}
\end{figure}
\begin{figure}[H]
\begin{center}
\includegraphics*[width=.8\textwidth]{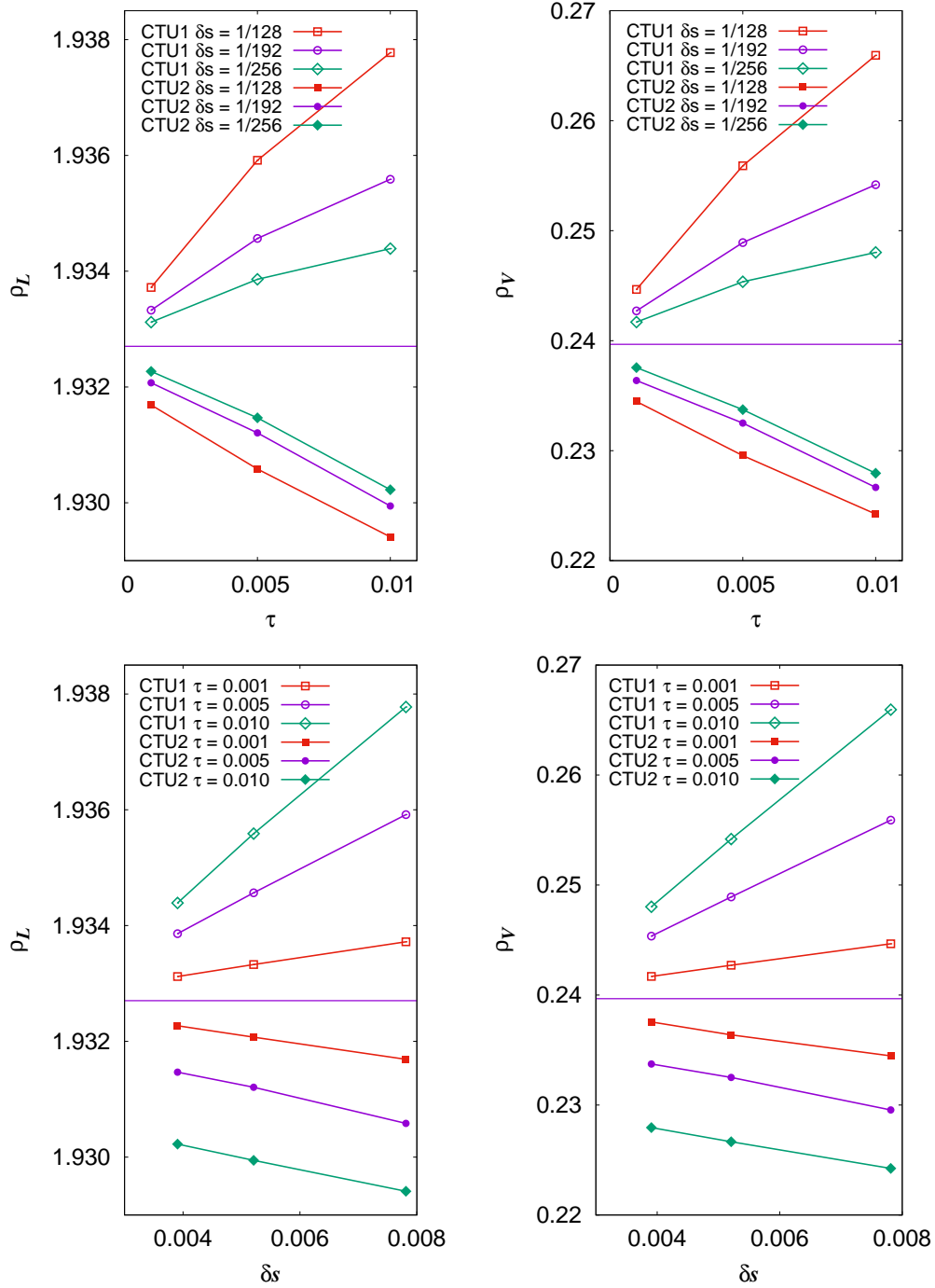} 
\end{center}
\caption{Dependence of the numerical
liquid and vapor densities $\rho_L$ and $\rho_V$, respectively,
on the relaxation
time $\tau$ and on the lattice spacing $\delta s$ at temperature $T=0.80$,
obtained with the CTU1 and the CTU2 numerical schemes
($\delta t = 10^{-4}$). The horizontal line
in each plot shows the corresponding theoretical density value computed using
the Maxwell construction.}
\label{fspacetau}
\end{figure}

The liquid-vapor phase diagram of the present model is shown
in Fig.~\ref{fphase}
was determined by inspecting the profile of the planar liquid-vapor interface
in the stationary case at various temperatures.
The simulations were conducted using the CTU2 
numerical scheme with the relaxation time $\tau=0.001$,
the time step $\delta t = 10^{-4}$, and the lattice spacing $\delta s = 1/256$.
Good agreement between the LB values of the liquid and vapor
densities and the corresponding values derived by the Maxwell construction
is seen for all temperatures $T \geq 0.70$. For lower temperatures, the
values of the vapor density become significantly smaller than the values
derived by the Maxwell construction (e.g., at T=0.60 , their relative difference
approaches $8\%$). As seen in Fig.~\ref{fspacetau},
when the relaxation time $\tau$ or the lattice spacing $\delta s$ decrease,
the values of both the liquid and the vapor densities approach the 
corresponding values derived using the Maxwell construction, regardless
of the numerical scheme (CTU1 or CTU2). This is not
a surprise if we recall that the LB simulation results approach the
results of the Navier-Stokes equations when the relaxation time
$\tau$ decreases
\cite{shan_jfm2006,ambrus_pre2012,ambrus_jcph2016,piaud_ijmpc2014,sspre2005,pre2014, sone,karniadakis} and, moreover,
the numerical errors induced by the finite volume schemes
always reduce when the lattice spacing decreases.


\section{Simulation results 
\label{cresults}}



\subsection{Critical radius for bubble growth in a quiescent liquid}
\label{snuma}


In this subsection, we will consider the kinetics of 
a vapor bubble expanding in a superheated liquid.
Let us denote by $\rho_{L}$ and $\rho_{V}$ the values of the
liquid and vapor densities of the van der Waals fluid, as calculated from the
non-dimensionalized equation of state \eqref{evdw}
according to the Maxwell construction.
When a vapor bubble of density $\rho_V$ and initial radius $R(t=0)$
is placed in a superheated liquid at density $\rho_{ext}<\rho_L$, 
it will shrink or grow depending on its initial
size since the system will tend to locally decrease its Gibbs free energy
density,
the latter being given by the Helmholtz free energy 
density $\psi$ plus the pressure.
Indeed, the system can reduce the Helmholtz free energy by increasing
the bubble size via phase separation of some of the metastable liquid to the
coexistence densities.
On the other hand this determines an increase
of the interfacial free energy as the bubble grows. The balance between
these two contributions, under the constraint of local mass conservation,
causes either the growth or the collapse of the bubble.
It has been shown \cite{laurila-2012}
that the critical radius $R_c$ of the bubble
that will neither shrink or grow is\footnote{We remark that Eq.~(41)
of Ref.~\cite{laurila-2012} contains a misprint since the exponent $-1$ 
on the r.h.s. is missing.}
\begin{equation}
R_c^{pred}=-\frac{\sigma}{2}\Big\{\Big[
\psi(\rho_V,T)-\psi(\rho_{ext},T)\Big]
+\frac{\rho_{ext}-\rho_V}{\rho_L-\rho_{ext}}\Big[\psi(\rho_L,T)
-\psi(\rho_{ext},T)
\Big]
\Big\}^{-1}
\label{eq:raggio_c}
\end{equation}
where $\sigma$ is the surface tension between liquid and vapor at coexistence and $\psi(\rho,T)$ is given by Eq.~\eqref{free-en2}.
The surface tension was numerically computed by using its definition
\begin{equation}
\sigma=\frac{\kappa}{2}\int dx \big[\nabla \rho(x)\big]^2
\label{eq:sigma}
\end{equation} 
where the numerical values of the density $\rho$ across a plane 
interface with liquid and vapor phases relaxed to equilibrium,
were used.

In order to test the prediction (\ref{eq:raggio_c}) in our model,
vapor bubbles at density $\rho_V = 0.2396$ with different values of the
initial radius $R(t=0)$ were centered in the lattice domain
and surrounded by a superheated liquid at density
$\rho_{ext} < \rho_L = 1.9327$.
The fluid density was allowed to evolve freely within
a circle of constant radius $R_{BC} = [L/2 - 1/(2\delta s)]\delta s$, 
where $L = L_x = L_y$ is the number of nodes on each Cartesian axis. 
Outside this circle, the liquid density
was set to the prescribed value $\rho_{ext}$ according to the
following procedure. At
time $t=n\delta t$, periodic boundary conditions were used to evolve the
distribution functions in all nodes of the lattice. Before processing the
next time step, the local fluid density $\rho_{i,j}^{n}$ was evaluated
in each lattice node $(i,j)$, $0\leq i,j < L$ and, if the node $(i,j)$
is located outside the circle of radius $R_{BC}$, the values of the 
corresponding distribution functions $f_{k;i,j}^{n}$
were rescaled by the factor $\rho_{ext} / \rho_{i,j}^{n}$.

In order to explore the effect of the lattice spacing on the accuracy of the
computer results, we conducted two series of computer simulations
with the CTU2 numerical  scheme. In the first series, we used a lattice
with $L = 2048$ nodes on each axis and spacing $\delta s = 1/128$, while
in the second series we used three lattices with
$L = 4096,\, 2048$ and $1024$ nodes, all with spacing $\delta s = 1/256$.
The other parameters of these runs were
$\delta t=10^{-4}$,  $\tau=10^{-3}$, $T=0.8$ and $\kappa=10^{-4}$.
The values of the surface tension $\sigma$ are quite independent on the lattice
spacing ($\sigma=4.8754\times 10^{-3}$ and
$\sigma=4.8747\times 10^{-3}$ for $\delta s = 1/128$ and $\delta s = 1/256$,
respectively).

\begin{figure}[H]
\begin{center}
\includegraphics*[width=.6\textwidth]{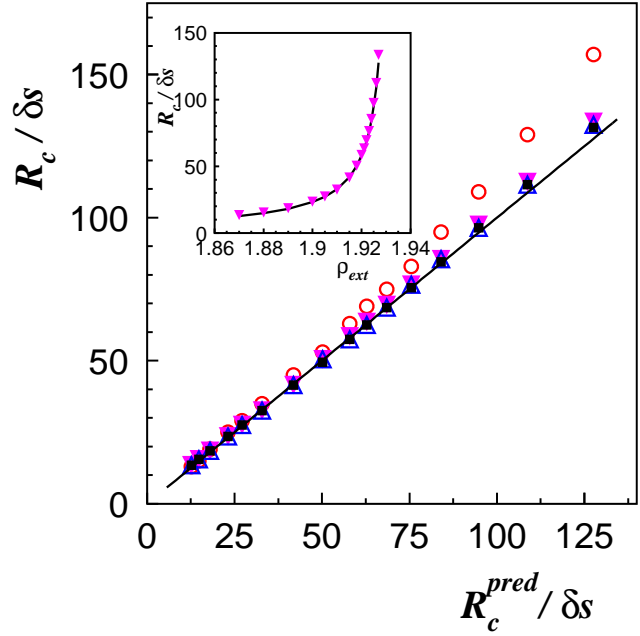}
\caption{Values of the critical bubble radius $R_c$ 
from LB simulations versus the theoretical predictions of 
Eq.~(\ref{eq:raggio_c}) represented by the full line.
The marks correspond to various values of the external density
$\rho_{ext}$. The LB results were obtained with the CTU1 scheme on a lattice with
spacing $\delta s = 1/128$ and size $L=2048$ ($\circ$) and
with the CTU2 scheme on lattices
with $\delta s = 1/256$  and sizes $L=4096 (\blacktriangledown),\,2048
(\triangle),\,1024 (\blacksquare)$. 
Inset: Values of $R_c$ 
from LB simulations with the CTU2 scheme on the lattice
with $\delta s = 1/256$  and size $L=4096(\blacktriangledown)$, and 
from  the theoretical predictions $(-\!\!\!-\!\!\!-)$ as a function
of the external density $\rho_{ext}$.}
\label{fig:r_crit}
\end{center}
\end{figure}

The evolution of the bubbles was monitored
for several values of $\rho_{ext}$ in the range $[1.870-- 1.927]$.
For each value of $\rho_{ext}$, the critical value $R_c$ of the bubble
 was estimated as $R_c=(R_g+R_s)/2$ where
$R_g$ is the initial smallest radius of a growing bubble and 
$R_s$ is the initial largest
radius of a shrinking bubble with $R_g=R_s+\delta s$. 
The numerical values of $R_c$ are plotted in Fig.~\ref {fig:r_crit},
where they are compared to the ones predicted by Eq.~(\ref{eq:raggio_c}). 
We note that Eq.~(\ref{eq:raggio_c}) predicts 
$R_c^{pred}$ to increase with $\rho_{ext}$ 
(see the inset of Fig.~\ref {fig:r_crit}). 
It appears that numerical results of $R_c$ agree
quite well with $R_c^{pred}$ for the smaller value of $\delta s$
with a slight overestimation at larger values of $\rho_{ext}$.
For this reason the rest of the study performed in this paper will be
done using the value $\delta s=1/256$ of the lattice spacing.
Finally, no dependence of the critical radius on the system size $L$ can be 
appreciated, as it appears from Eq.~(\ref{eq:raggio_c}). This is quite
well confirmed
in Fig.~\ref {fig:r_crit}, when comparing the corresponding
values of $R_c$ obtained on the three lattices with $\delta s = 1/256$.


\subsection{Bubble growth in a quiescent liquid: The Rayleigh-Plesset
equation}\label{srp}

\begin{figure}[H]
\begin{center}
\includegraphics*[width=.5\textwidth]{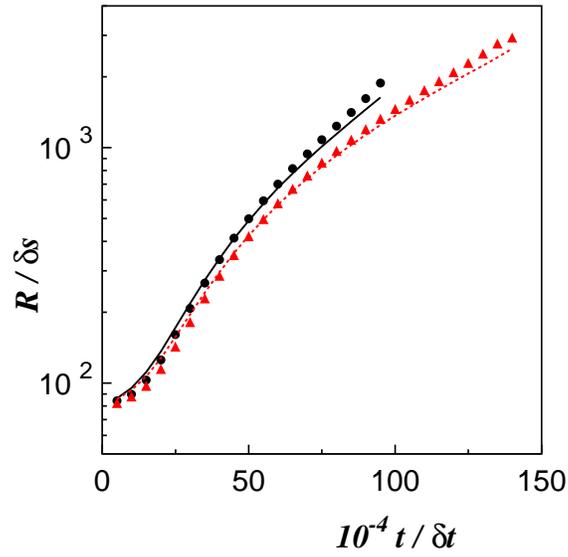}
\caption{The radius $R$ of the growing bubble in a quiescent liquid
as a function
of time for lattice size $L=4096 (\bullet), 6144 (\blacktriangle)$ from the
lattice Boltzmann simulations. The full and dashed lines correspond
to the numerical solutions of the RP equation (\ref{eq:rp})
for $L=4096, 6144$, respectively.}
\label{fig:r_rp}
\end{center}
\end{figure}

As we saw above, a vapor bubble
immersed in a superheated liquid at density $\rho_{ext} < \rho_L$
will grow when its initial radius is larger than the corresponding
critical value $R_c$. For some values of $\rho_{ext}$, we followed the
evolution of the radius  $R(t)$ of vapor bubbles
of initial size $R > R_c$ and density $\rho_V = 0.2396$
 on lattices of size $L=4096, 6144$,
with $\delta s = 1/256$ and density fixed at the value $\rho_{ext}$ 
at the nodes outside the circle of radius $R_{BC}$, as already described
in the previous section.
The evolution of the bubble radius was followed after the relaxation
of the initial sharp interface. The bubble keeps a circular shape
during the overall process.
The results of $R(t)$ versus time
shown in Fig.~\ref{fig:r_rp} were obtained for an initial bubble radius
$R =  77\, \delta s$, which is larger than the value $R_c = 76.5\,\delta s$
corresponding to the choice $\rho_{ext}=1.923$.
Results for other values of $\rho_{ext}$ are similar.
Before commenting the results, we discuss the equation which
describes the evolution of the bubble radius for the present problem.

The time behavior of the radius of a spherical vapor bubble in an infinitely
large liquid domain at constant temperature is described by 
the Rayleigh - Plesset (RP) equation \cite{brennen-1995}. 
In the following we will derive for completeness
its form in the two-dimensional case\footnote{The
expressions previously reported in Refs.~\cite{chen-2010,chen-2011} 
contain some misprints.}.
We consider a circular vapor bubble of radius $R$
in a liquid whose density $\rho_L$ and dynamic viscosity $\mu_L$
are assumed constant.
The radial position will be denoted by the distance $r$ from
the bubble center ($r=0$) located in the middle
of the system, the pressure by $p(r,t)$, and
the radial outward velocity by $u(r,t)$. The tangential component
of the velocity is null since the system has central symmetry.
The liquid far field boundary
is located at $r_{\infty}=R_{BC}$,  where the pressure is $p_{\infty}$.
The pressure $p_B$ and the density $\rho_{V,B}$ inside the bubble
are assumed to be uniform. In order to guarantee mass conservation it is
taken
\begin{equation}
u(r,t)=\frac{F(t)}{r}
\label{u}
\end{equation}
where $F(t)$ is a function to be determined
in order to satisfy the continuity equation which for an 
incompressible fluid reads as
\begin{equation}
\frac{1}{r} \frac{\partial}{\partial r}  \big [ r u(r,t) \big ]=0.
\end{equation}
$F(t)$ and $R(t)$ are related by a kinematic boundary condition at 
the bubble interface.
Assuming that there is no mass flow across this interface, it has
to be $u(R,t)=dR/dt$ and hence 
\begin{equation}
F(t)=R \frac{dR}{dt} .
\label{f}
\end{equation}
Equation (\ref{f}) holds also in the presence of evaporation or
condensation at the interface under the hypothesis that 
$\rho_L >> \rho_{V,B}$
\cite{brennen-1995}.

In the case of a Newtonian liquid, the Navier-Stokes
equation for the radial velocity is 
\begin{equation}
\frac{\partial u(r,t)}{\partial t}
+ u(r,t) \frac{\partial u(r,t)}{\partial r}=
-\frac{1}{\rho_L} \frac{\partial p(r,t)}{\partial r}+
\frac{\mu_L}{\rho_L}\Big [ 
\frac{1}{r}\frac{\partial}{\partial r}
\big (r \frac{\partial u(r,t)}{\partial r} 
\big )
- \frac{u(r,t)}{r^2} 
\Big ] .
\label{ns}
\end{equation}
Substituting Eq.~(\ref{u}) into Eq.~(\ref{ns}) and
then integrating from $R$ to $r_{\infty}$ yields
\begin{equation}
\ln \Big(\frac{r_{\infty}}{R}\Big) \frac{dF(t)}{dt}-\frac{F^2(t)}{2}
\Big(\frac{1}{R^2}-\frac{1}{r_{\infty}^2}\Big)=
\frac{p(R)-p_{\infty}}{\rho_L} .
\label{ns2}
\end{equation}
Moreover, a pressure boundary condition on the interface can be introduced 
which is obtained by fixing to zero the total force per unit length
on the interface in the absence of mass transport across the boundary
\cite{brennen-1995}:
\begin{equation}
p(R)=p_B-\frac{\sigma}{R}-\frac{2\mu_L}{R} \frac{dR}{dt} .
\label{bc}
\end{equation}
Substituting Eqs.~(\ref{f}) and (\ref{bc}) into Eq.~(\ref{ns2}) delivers
the final form of the two-dimensional RP equation 
\begin{equation}
\ln \left(\frac{r_{\infty}}{R}\right) \left[ \left(\frac{dR}{dt}\right)^2 
+ R \frac{d^2R}{dt^2}\right]
-\frac{1}{2}\left[1-\frac{R^2}{r_{\infty}^2}\right] \left(\frac{dR}{dt}\right)^2
+ \frac{\sigma}{\rho_L R}+ \frac{2 \mu_L}{\rho_L R} \frac{dR}{dt}
= \frac{p_B(t)-p_{\infty}(t)}{\rho_L} .
\label{eq:rp}
\end{equation}

Some comments are here in order about Eq.~(\ref{eq:rp}). 
It is evident that the growth of the
bubble radius $R$ depends on the spatial extension 
$r_{\infty}$ of the
system differently from the three-dimensional case.
This is due to
the $1/r$ dependence of $u(r,t)$ in Eq.~(\ref{u}) which gives rise
to the logarithmic term in RP equation. Once $p_{\infty}(t)$ is given,
RP equation can be solved to find $R(t)$ if $p_B(t)$ is known.
We solved it numerically by using a Runge-Kutta method to compare
the results to the output of LB simulations.
To this purpose
the values of $\rho_L$, $\mu_L$, and $\sigma$ are the ones of the present
LB model. Moreover, $p_B(t)$ and $p_{\infty}(t)$
were measured plugging into the EOS the values of density
at the bubble center ($r=0$) and at the 
domain boundary  ($r=r_{\infty}$), respectively,
obtained from the LB simulations. The initial values of $R$
and $dR/dt$ were taken from the LB runs after the initial
relaxation of the bubble interface.

The evolution of the bubble radius $R(t)$ is plotted 
in Fig.~\ref{fig:r_rp} for both the numerical solutions of RP equation and
the LB simulations on lattices with $L\times L$ nodes. 
Since $r_{\infty}/\delta s=L/2-1/(2\delta s)$ is a finite quantity in our
model, the results of the LB simulations are expected to depend
on the lattice size $L$.  Indeed, although both the LB results reported in 
Fig.~\ref{fig:r_rp} are quite in good agreement to the 
numerical solutions of the RP equation only in the early stage of the
bubble growth process, the results on the smaller lattice start 
to deviate from the predictions of the RP equation at time
$t_{4096}\simeq 7 \times 10^5 \delta t$,
while the results on the larger lattice are still consistent up to  
$t_{6144} \simeq 10^6 \delta t$. This is due to the fact that
the RP equation 
relies on the implicit assumption of an infinite liquid domain
where the ratio $r_{\infty}/R(t)$ is very large. When this ratio is small,
the first two terms in the RP equation \eqref{eq:rp} become negligible
and the RP equation loses its meaning.
In the case of the larger lattice, this ratio is 
$r_{\infty} / R(t_{6144}) \simeq 2$
and continues to reduce at times $t > t_{6144}$, worsening
the agreement between the LB simulation results and the RP equation.
The present model is thus capable to account for
the bubble growth
up to a lattice-size dependent time $t_L$,
while remaining in good agreement to the RP
equation  until $R(t_L)/R(0) \simeq 20$. 
This value is considerably larger than the one 
($\simeq 5$) reached in previous studies 
\cite{chen-2010,chen-2011}. 


\subsection{Bubble growth under shear flow}\label{sshear}


The behavior of an equilibrated vapor bubble of density $\rho_V$ 
and dynamic viscosity $\mu_V$ with radius
$R$
in a liquid with density $\rho_L$ and dynamic viscosity $\mu_L$
under shear flow received 
considerable attention in the past 
\cite{acrivos-stone,rallison-1984,brennen-1995}.
Here we will briefly sketch the phenomenology. For weak flows such that
the capillary number $\displaystyle Ca=\frac{\mu_L \dot\gamma R}{\sigma} <<1$,
$\dot\gamma$ being the shear rate, 
the bubble is deformed assuming in the stationary regime
an elliptical shape whose principal axis forms a tilt angle 
$\theta \simeq \pi/4$ with the flow direction. 
When increasing the shear rate, the
equilibrium shape of the bubble is more elongated with $\theta$ decreasing
to zero independently on the value of the viscosity
ratio $\lambda=\mu_V/\mu_L$. A further
increase of the shear rate would deform the bubble into a point-ended shape
until its break-up at small
values of $\lambda$, while for $\lambda > \lambda_c \simeq 4$ 
the bubble would attain an equilibrium elliptical
shape with $\theta \simeq 0$. 

In the present study a lattice of size $L \times L$ with $L=6144$
was confined by two permeable horizontal walls shearing with velocities
${\mathbf{u}}_{top} = (u_w,0)$ and ${\mathbf{u}}_{bot} = (-u_w,0)$
along the $x$ axis, respectively. In the lattice nodes
outside the walls, i.e., in the ghost nodes
$(i,j)$, $0 \leq i < L$, $j \in \{-2,\,-1,\,L,\,L+1\}$,
the distribution functions $f_{k;i,j}^{t}$ were set according to 
Eq.~\eqref{herfeq}, where $\rho$ was replaced by $\rho_{ext}$ and
\begin{equation}
{\mathbf{u}} = \left\{
\begin{array}{rcl}
{\mathbf{u}}_{top} & , & j \in \{L,\,L+1\} \\ 
{\mathbf{u}}_{bot} & , & j \in \{-2,\,-1\}  
\end{array}
\right.
\end{equation}
Periodic boundary conditions were applied in the
horizontal direction.

A bubble of initial radius $R=26 \delta s$ and density
$\rho_V = 0.23967$ was placed in a superheated liquid with density
$\rho_{ext}=1.90$ at $T=0.80$. Under these conditions, the bubble grows in
a quiescent liquid as 
previously seen. Various values of the wall velocity $u_w$
were considered in order to vary
the shear rate $\dot\gamma=2 u_w/(L \delta s)$. The highest value of
$u_w$ was such to have Mach number $Ma=u_w/c_s \simeq 0.5$,
where $c_s=\sqrt{dp^{w}/d\rho} \simeq 1.7$ is the sound velocity in the
liquid phase.
We remark that the present model, being accurate at the third order
with the correct quadrature,
is not limited to the incompressible regime \cite{chen-2008}.
Because of the large system size here adopted to follow the growth of the
bubble on long time scales, 
the values of $\dot\gamma$ are small so we
considered the relaxation time $\tau=10^{-2}$
in order to increase the liquid viscosity
$\mu_L$ ($=\rho_L T \tau$) and, thus, accessing larger values of the
capillary number.
The fluid velocity was initialized to be the one corresponding to a
linear flow profile with shear rate $\dot\gamma$. The bubble 
grew by the same mechanism previously described being, in the meanwhile,
deformed and rotated by shear. 

The morphology and alignment with the flow were studied by
using the gyration tensor of the bubble,
defined as
\begin{equation}
T_{\alpha\beta} = \frac{1}{N_b} \sum_{i \in bubble}
 (r_{i,\alpha} - {\bar{r}_{\alpha}})  (r_{i,\beta} - {\bar{r}_{\beta}})
\label{gyr}
\end{equation}
where the sum is over the $N_b$ lattice sites belonging to the bubble,
whose position vectors are ${\bm{r}}_{i}$.
The position vector of the center of the bubble is
${\bm{\bar{r}}} = \sum_{i \in bubble} {\bm{r}}_{i} / N_b$.
The two eigenvalues $\Lambda_M$ and $\Lambda_m$ with $\Lambda_M >
\Lambda_m$ of the gyration tensor were then used to characterize
the bubble shape. Indeed, in the case of an ellipse with semi-axes
$a$ and $b$ with $a>b$ it can be shown that $a=2 \sqrt{\Lambda_M}$ and
$b=2 \sqrt{\Lambda_m}$. This will be the way here adopted to estimate
the typical size of the elliptical bubble.
However, we checked that
the results later presented do not depend on this particular way of estimating
$a$ and $b$. Indeed, for a comparison $a$ and $b$ were also computed as
the largest and smallest distances from the bubble center to the interface
located at density $\rho = (\rho_V+\rho_L)/2$, 
respectively, finding no difference. 
Since the bubble is deformed while $a$ and $b$
grow in time (see the next discussion), the average size of the bubble
is defined as $\hat R=(a+b)/2$,
which depends on time via $a$ and $b$.
Consequently the capillary number
is now computed as
$Ca=\mu_L \dot\gamma \hat R/\sigma$ and depends on time.
The deformation of the bubble
is expressed in terms of the dimensionless number
$D=(a-b)/(a+b)$ \cite{taylor-1932}. Finally, 
the tilt angle $\theta$ of the bubble is computed by measuring the angle formed
by the eigenvector of $T_{xy}$ 
corresponding to $\Lambda_M$ with the flow direction.

\begin{figure}[H]
\begin{center}
\includegraphics*[width=.5\textwidth]{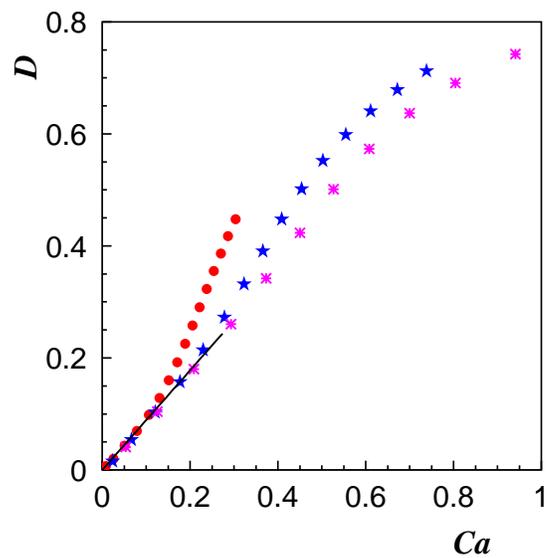}
\caption{The deformation $D$  of the bubble as a function
of the capillary number $Ca$ in a lattice of size  $L=6144$
for shear rates $\dot \gamma \delta t=1.67 
\times 10^{-6} (\bullet), 3.33 \times 10^{-6} (\star), 5.00 \times 10^{-6}
(\ast)$. The full line has slope $0.89$.}
\label{fig:deform}
\end{center}
\end{figure}

\begin{figure}[H]
\begin{center}
\includegraphics*[width=.9\textwidth]{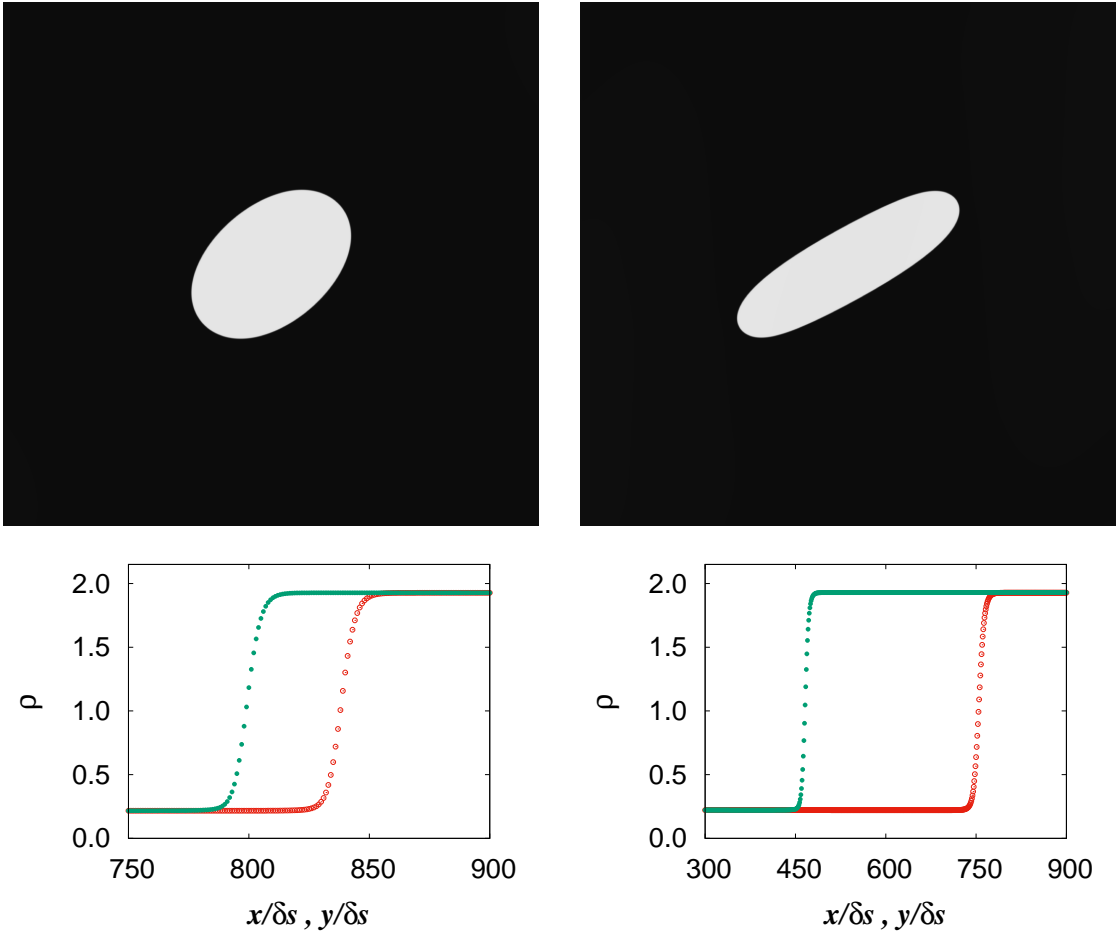}
\caption{Density plots of the bubble (upper panels)
at time $t/\delta t= 5 \times 10^5$  
in a lattice of size  $L=6144$ and the
density profiles (lower panels) in the interface regions, plotted
along the Cartesian axes $x (\bullet)$ and $y (\circ)$) 
centered in the middle of the flow domains,
for shear rates
$\dot \gamma \delta t=1.67 \times 10^{-6}$ (left) and 
$5.00 \times 10^{-6}$ (right).
The values of the capillary number are $Ca=0.18$ (left )and $0.61$ (right).
}
\label{fig:dens0106}
\end{center}
\end{figure}

The behavior of $D$ as a function of $Ca$ is shown in
Fig.~\ref{fig:deform} for various
values of the shear rate. Simulations are run until the bubble reaches
the boundary.
It can be seen that the deformation increases linearly with the capillary
number up to $Ca \simeq 0.2$ 
and is independent on the value
of the shear rate as previously observed \cite{chen-2010,chen-2011}.
This can be compared with the prediction in the case
of an equilibrated bubble under steady deformation for weak flows where
it holds that $D=(19\lambda+16) Ca / (16 \lambda+16)$ for 
$Ca << 1$ \cite{taylor-1932}. This would give $D \simeq 1.02 Ca$
for the value $\lambda \simeq 0.12$ of our system.
The best
fit to numerical data gives $D \simeq 0.89 Ca$. 
We stress that in our case the relationship between $D$ and $Ca$ is dynamic
in the sense that both quantities depend on time keeping the
shear rate fixed, while in the
case of steady deformation $D$ is obtained by considering successive
increments of $Ca$ by increasing the shear rate. 
When the capillary number further increases beyond $0.2$,
the deformation is no longer 
a linear function of $Ca$ and the smaller is the shear rate 
the higher is the deformation with no 
overlap of data for the different values of the shear rate.
One expects that high order contributions
of $Ca$ to $D$ might be relevant also in the present problem
as it is in the case of steady 
deformation \cite{barthes-1973}. 
Typical bubble conformations in the two regimes are  
shown in Fig.~\ref{fig:dens0106} at the same time 
for $\dot \gamma \delta t=1.67 \times 10^{-6}, 5.00 \times 10^{-6}$.
For the lower value of $\dot\gamma$ it results $Ca=0.18$ so that
the deformation is still linear in $Ca$
while in the other case it is
$Ca=0.61$ when $D$ is no longer a linear function of the
capillary number (see Fig.~\ref{fig:deform}).
In the same figure the finite width of the bubble interfaces
along the flow and the shear directions can be appreciated
with no deformation induced by the external flow.
We are able to observe 
a non-linear regime of $D$ as a function of $Ca$ 
in the case of a sheared growing bubble thanks to the very large
simulated system.

\begin{figure}[H]
\begin{center}
\includegraphics*[width=.5\textwidth]{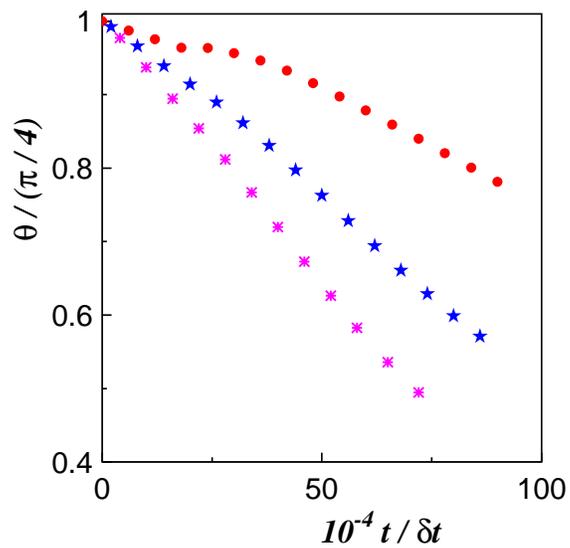}
\caption{The tilt angle $\theta$   of the bubble as a function
of time in a lattice of size  $L=6144$
for shear rates $\dot \gamma \delta t=1.67 
\times 10^{-6} (\bullet), 3.33 \times 10^{-6} (\star), 5.00 \times 10^{-6}
(\ast)$.}
\label{fig:angle}
\end{center}
\end{figure}

The time behavior of the tilt angle $\theta$, which is reported in
Fig.~\ref{fig:angle}, is observed to depend on shear rate. 
At the beginning the elongational component of the shear flow aligns
the slightly deformed bubble along the direction of principal extension 
so that $\theta \simeq \pi/4$. Immediately afterward the angle diminishes
and the lower is the shear rate, the higher is the tilt angle with
a linear dependence of $\theta$ on the shear rate. However,
at late times this dependence is no longer linear.

\begin{figure}[H]
\begin{center}
\includegraphics*[width=.5\textwidth]{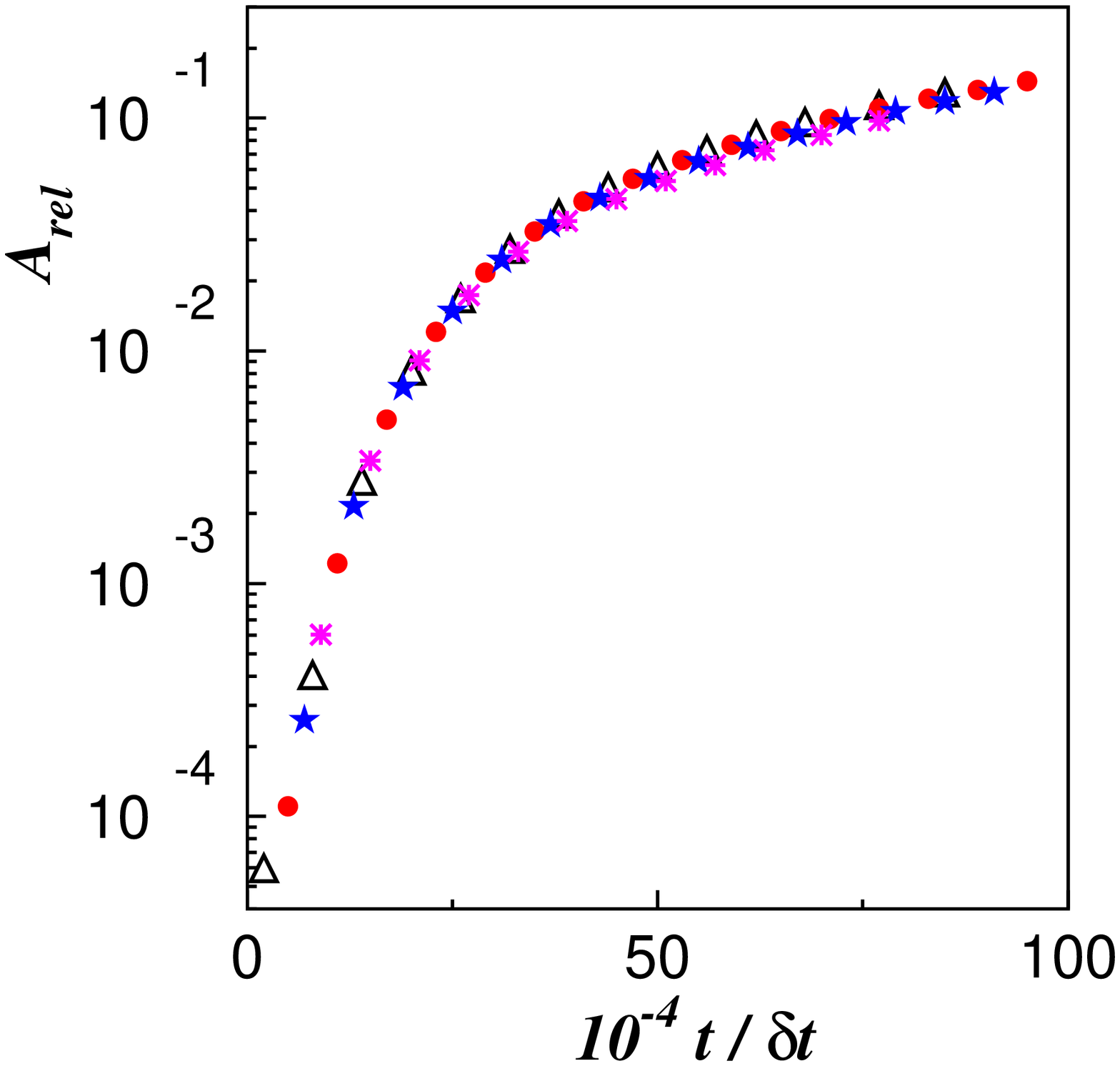}
\caption{The fraction $A_{rel}$ of the bubble area as a function
of time in a lattice of size  $L=6144$
for shear rates $\dot \gamma \delta t=0 (\triangle),
1.67 \times 10^{-6} (\bullet), 3.33 \times 10^{-6} (\star), 5.00 \times 10^{-6}
(\ast)$.}
\label{fig:area}
\end{center}
\end{figure}

In order to evaluate the effects of the shear on the growth rate 
of the bubble, the fraction $A_{rel}=N_b/L^2$ of the bubble area with respect
to the system
extension was computed. 
The results as a function of time are depicted
in Fig.~\ref{fig:area}. 
It can be appreciated that the area of the bubble
does not depend on the shear rate, even with steady walls, 
showing that the growth is mainly
driven by the pressure difference. 

Finally, we comment about the possibility of accessing larger values
of the capillary number. Within the present model it is hard
to go beyond $Ca \simeq 1$. Indeed, it can be noted that
$Ca=\mu_L \dot\gamma \hat R/\sigma \simeq Ma \rho_L \tau T^{3/2}/\sigma$.
The numerator cannot be further increased with respect to the present
study since $Ma \simeq 0.5$, $\tau \lesssim 10^{-2}$ due to the 
constraint on the validity of the Navier-Stokes limit, and
$T < T_c=1$. The only way would be to diminish the surface tension.
Since it can be shown \cite{wagn07} that 
$\sigma \simeq (\rho_L-\rho_V)^2 \sqrt{\kappa(1/T-1)/2}$,
one might reduce $\kappa$ and/or increase $T$ with $T<1$. However,
since the interface width is proportional to
$\sqrt{2\kappa/(1/T-1)}$ \cite{wagn07}, a reduction of $\kappa$ would
make the interface sharper compromising the numerical stability of the method
and an increase of $T$ would broaden the interface requiring larger systems
to keep the same resolution thus making the simulation not feasible.


\section{Conclusion}


\label{sec:conclusion}
We  introduced a third-order, off lattice isothermal LB model in two dimensions 
with the purpose to  describe 
the growth behavior of a vapor bubble in superheated liquid.
The model is based on the Gauss-Hermite quadrature and 
on the second-order corner transport upwind numerical scheme which is
easily parallelizable as the collision-streaming LB models.

We first considered  a quiescent system.  
We presented a corrected version of the two-dimensional
Rayleigh-Plesset equation and
found that our numerical results  well describe
the evolution of the radius of the bubble. The agreement with the solution
of RP equation becomes better 
for larger sizes of the system. We remind that, differently from the three-
dimensional case, 
the spatial extension of the system explicitly enters in the formulation of the RP 
equation in two dimensions. We also presented a careful
evaluation of the critical radius of a bubble for the non-equilibrium conditions considered in this work. 

Then we analyzed the same problem in presence of a shear flow imposed by 
external walls.  We measured the growth and the deformation of the bubble
induced by the flow. We expressed the deformation in terms of the 
dimensionless number $D$ and analyzed its dependence on
the capillary number $Ca$ that is evaluated in terms of shear rate and 
average radius of the bubble. 
As expected, a linear dependence was observed at low $Ca$ but with a 
different proportionality coefficient
than that known for bubbles in equilibrium liquids.  This coefficient was found to 
be the same for the different
shear rates considered. A non-linear regime was observed for $Ca \gtrsim 0.2$ 
with $D$ being slightly larger, at fixed $Ca$, for smaller shear rates. 
In a future research we plan to extend our method and analysis in order to 
control independently the viscosities
of the liquid and vapor phases.


\acknowledgments


V.S., T.B., S.B. and V.E.A. are supported by a grant of the
Romanian National Authority for Scientific Research,
CNCS-UEFISCDI, Project No. PN-II-ID-PCE-2011-3-0516.
V.E.A. gratefully acknowledges the support of NVIDIA Corporation 
with the donation of a Tesla K40 GPU used for this research.
G.G. acknowledges partial support from MIUR, Project  
PON 02-00576-3333604  INNOVHEAD.


\appendix*\section{}\label{app}

\renewcommand{\theequation}{A.\arabic{equation}}


In order to clarify what happens with the series expansion \eqref{exfh2}
during the advection step, we first recall the definition of the tensor
Hermite polynomials
in the $D$-dimensional Cartesian space \cite{ shan_jfm2006} :
\begin{equation}
{\bm{\mathcal{H}}}^{(\ell)}_{\alpha_{1}\ldots\alpha_{\ell}}({\bm{\xi}}) \, = \,
\frac{\,(-1)^{\ell}\,}{\,\omega({\bm{\xi}})\,} \,
\partial_{\xi_{\alpha_{1}}} \cdots
\partial_{\xi_{\alpha_{\ell}}} \omega({\bm{\xi}})
\end{equation}
where $\ell \in \{\, 0,\,1,\,2,\,\ldots \,\}$,
 $\alpha_{1},\,\alpha_{2},\,\ldots \alpha_{\ell}\,\in
\{\, x_{1},\, x_{2},\, \ldots\, x_{D}\,\}$ and
\begin{equation}
\omega({\bm{\xi}}) \,=\, \frac{1}{\,(2\pi)^{D/2}\,}\,\exp(-\xi^{2}/2)
\,=\,\prod_{k=1}^{D} \frac{1}{\,\sqrt{2\pi}\,}\,\exp(-\xi_{x_{k}}^{2}/2)
\end{equation}
with
\begin{equation}
\xi^{2} \,=\, \sum_{k=1}^{D} \xi_{k}^{2} .
\end{equation}
The Hermite polynomials $H^{m}(\xi_{\alpha})$ of order $m$,
$m \in \{\,0,\,1,\,2,\,\ldots\,\}$, are defined on the Cartesian axis
$\alpha$, in a similar way :
\begin{equation}
H^{m}(\xi_{\alpha}) \,=\, \frac{\,(-1)^{m}\,}{\,\omega(\xi_{\alpha})\,} \,
\partial_{\xi_{\alpha}}^{m}  \omega(\xi_{\alpha})
\end{equation}
and satisfy the recurrence relation
\begin{equation}
\xi_{\alpha} H^{(m)} (\xi_{\alpha})  \,=\,
H^{(m+1)}(\xi_{\alpha})  \,+\,
m\, H^{(m-1)}(\xi_{\alpha})
\label{e6}
\end{equation}

In the two-dimensional space, we have $D=2$ and $x_{1} \equiv x$ ,
$x_{2} \equiv y$. We use the Kronecker symbol
$\delta_{\alpha\beta}$ to write:
 \begin{equation}
\partial_{\xi_{\alpha}} \,=\, \delta_{\alpha x}\partial_{\xi_{x}} +
\delta_{\alpha y}\partial_{\xi_{y}} .
\end{equation}
This allows us to express the tensor Hermite polynomials
${\bm{\mathcal{H}}}^{(\ell)}_{\alpha_{1}\ldots\alpha_{\ell}}({\bm{\xi}}) $
with respect to the Hermite polynomials $H^{m}(\xi_{\alpha})$ : 
\begin{eqnarray}
{\bm{\mathcal{H}}}^{(\ell)}_{\alpha_{1}\ldots\alpha_{\ell}}({\bm{\xi}}) & = &
\frac{\,(-1)^{\ell}\,}{\,\omega({\bm{\xi}})\,} \,
\prod_{k=1}^{\ell}
(\delta_{\alpha_{k} x}\partial_{\xi_{x}} +
\delta_{\alpha_{k} y}\partial_{\xi_{y}})
\, \omega({\bm{\xi}}) \nonumber \\
& = & 
\sum_{{\tiny{\begin{array}{c}
m,n=0 \\ m+n=\ell \end{array}}}
}^{\ell} \,\delta_{(m,n)}^{(\ell)}\,
\left[\,
\frac{\,(-1)^{m}\,}{\,\omega(\xi_{x})\,}\, \partial_{\xi_{x}}\omega(\xi_{x})
\,\right] \, \left[
\frac{\,(-1)^{n}\,}{\,\omega(\xi_{y})\,} \, \partial{\xi_{y}}\omega(\xi_{y})\,
\right] \nonumber\\
& = &
\sum_{{\tiny{\begin{array}{c} m,n=0 \\ m+n=\ell \end{array}}}}^{\ell} \,
\delta_{(m,n)}^{(\ell)}\, H^{m}(\xi_{x}) H^{n}(\xi_{y})
\label{lmn}
\end{eqnarray}
where the symbol $\delta_{(m,n)}^{(\ell)}$, with $m+n=\ell$, 
is defined recursively, as follows. For $\ell=m=n=0$, we set
\begin{equation}
\delta_{(0,0)}^{(0)} \, = \, 1 .
\end{equation}
For $\ell>0$, when $m=\ell$ or $n=\ell$, we define
\begin{eqnarray}
\delta_{(\ell,0)}^{(\ell)} & = & \delta_{(\ell-1,0)}^{(\ell-1)} \delta_{\alpha_{l}x}\\
\delta_{(0,\ell)}^{(\ell)} & = & \delta_{(0,\ell-1)}^{(\ell-1)} \delta_{\alpha_{l}y}
\rule{0mm}{7mm}
\end{eqnarray}
and, for $\ell,m,n>0$, $m+n=\ell$ :
\begin{equation}
\delta_{(m,n)}^{(\ell)} \, = \, \delta_{(m-1,n)}^{(\ell-1)} \delta_{\alpha_{l}x} \,+\,
\delta_{(m,n-1)}^{(\ell-1)} \delta_{\alpha_{l}y} .
\end{equation}

In this way we are able to get the expansion of the tensor Hermite
polynomials up to order $N=4$ with respect to the Hermite polynomials:
\begin{eqnarray}
{\bm{\mathcal{H}}}^{(0)}({\bm{\xi}}) & = & H^{(0)}(\xi_{x}) H^{(0)}(\xi_{y}) ,
\label{he0} \\
{\bm{\mathcal{H}}}^{(1)}_{\alpha}({\bm{\xi}}) & =  &
\delta_{\alpha x} H^{(1)}(\xi_{x}) H^{(0)}(\xi_{y}) \, + \,
\delta_{\alpha y} H^{(0)}(\xi_{x}) H^{(1)}(\xi_{y}) , \rule{0mm}{7mm}\\
{\bm{\mathcal{H}}}^{(2)}_{\alpha\beta}({\bm{\xi}})
 & = &
 \delta_{\alpha x} \delta_{\beta x}  H^{(2)}(\xi_{x}) H^{(0)}(\xi_{y}) 
\nonumber \rule{0mm}{7mm} \\
 & + &
(\delta_{\alpha x} \delta_{\beta y} \,+\, 
\delta_{\alpha y} \delta_{\beta x} )
 H^{(1)}(\xi_{x}) H^{(1)}(\xi_{y})  \rule{0mm}{7mm}
 \nonumber \\  & + &
\delta_{\alpha y} \delta_{\beta y}  H^{(0)}(\xi_{x}) H^{(2)}(\xi_{y}) ,
 \rule{0mm}{7mm} \\
{\bm{\mathcal{H}}}^{(3)}_{\alpha\beta\gamma}({\bm{\xi}}) & = &
\delta_{\alpha x} \delta_{\beta x}\delta_{\gamma x} 
 H^{(3)}(\xi_{x}) H^{(0)}(\xi_{y})
  \nonumber \rule{0mm}{7mm} \\
& + &
(\delta_{\alpha x} \delta_{\beta x}\delta_{\gamma y} \,+\,
\delta_{\alpha x} \delta_{\beta y}\delta_{\gamma x} \,+\,
\delta_{\alpha y} \delta_{\beta x}\delta_{\gamma x})\,
 H^{(2)}(\xi_{x}) H^{(1)}(\xi_{y})
 \nonumber \rule{0mm}{7mm} \\
& + &
(\delta_{\alpha y} \delta_{\beta y}\delta_{\gamma x} \,+\,
\delta_{\alpha y} \delta_{\beta x}\delta_{\gamma y} \,+\,
\delta_{\alpha x} \delta_{\beta y}\delta_{\gamma y})\,
 H^{(1)}(\xi_{x}) H^{(2)}(\xi_{y})
 \nonumber \rule{0mm}{7mm} \\
& + &
\delta_{\alpha y} \delta_{\beta y}\delta_{\gamma y}\,
 H^{(0)}(\xi_{x}) H^{(3)}(\xi_{y})
\rule{0mm}{7mm} ,\\ 
{\bm{\mathcal{H}}}^{(4)}_{\alpha\beta\gamma\sigma}({\bm{\xi}}) & = &
\delta_{\alpha x} \delta_{\beta x}\delta_{\gamma x}\delta_{\sigma x}\,
 H^{(4)}(\xi_{x}) H^{(0)}(\xi_{y}) 
  \nonumber \rule{0mm}{7mm} \\
  & + &
(\delta_{\alpha x} \delta_{\beta x}\delta_{\gamma x}\delta_{\sigma y} \,+\,
 \delta_{\alpha x} \delta_{\beta x}\delta_{\gamma y}\delta_{\sigma x} \,+\,
 \delta_{\alpha x} \delta_{\beta y}\delta_{\gamma x}\delta_{\sigma x} \,+\,
 \rule{0mm}{7mm} \nonumber\\
&  &
\,\, \delta_{\alpha y} \delta_{\beta x}\delta_{\gamma x}\delta_{\sigma x})\,
   H^{(3)}(\xi_{x}) H^{(1)}(\xi_{y})
   \rule{0mm}{7mm} \nonumber\\
   & + &
(\delta_{\alpha x} \delta_{\beta x}\delta_{\gamma y}\delta_{\sigma y} \,+\,
 \delta_{\alpha x} \delta_{\beta y}\delta_{\gamma x}\delta_{\sigma y} \,+\,
 \delta_{\alpha y} \delta_{\beta x}\delta_{\gamma x}\delta_{\sigma y} \,+\,
 \rule{0mm}{7mm} \nonumber\\
&  & \,\,
\delta_{\alpha x} \delta_{\beta y}\delta_{\gamma y}\delta_{\sigma x} \,+\,
\delta_{\alpha y} \delta_{\beta x}\delta_{\gamma y}\delta_{\sigma x} \,+\, \delta_{\alpha y} \delta_{\beta y}\delta_{\gamma x}\delta_{\sigma x}) \,
 H^{(2)}(\xi_{x}) H^{(2)}(\xi_{y})
   \rule{0mm}{7mm} \nonumber\\
     & + &
(\delta_{\alpha y} \delta_{\beta y}\delta_{\gamma y}\delta_{\sigma x} \,+\,
 \delta_{\alpha y} \delta_{\beta y}\delta_{\gamma x}\delta_{\sigma y} \,+\,
 \delta_{\alpha y} \delta_{\beta x}\delta_{\gamma y}\delta_{\sigma y} \,+\,
 \rule{0mm}{7mm} \nonumber\\
&  &
\,\, \delta_{\alpha x} \delta_{\beta y}\delta_{\gamma y}\delta_{\sigma y})\,
   H^{(0)}(\xi_{x}) H^{(3)}(\xi_{y})
   \rule{0mm}{7mm} \nonumber\\
   & + &
   \delta_{\alpha y} \delta_{\beta y}\delta_{\gamma y}\delta_{\sigma y}\,
 H^{(0)}(\xi_{x}) H^{(4)}(\xi_{y}) .
     \rule{0mm}{7mm} \label{he4}
 \end{eqnarray}

\begin{table}
\caption{$(m,n)$ pairs in the expansion \eqref{lmn} of the function $f_{k}$ (see the text for further details).}
\label{t1}
\begin{center}
\begin{tabular}{ccccccccc}
{\textcolor{red}{[0,4]}} & \quad &
{\textcolor{blue}{[1,4]}} & \quad &
{\textcolor{green}{[2,4]}} & \quad &
{\textcolor{cyan}{[3,4]}} & \quad &
{\textcolor{magenta}{[4,4]}} 
\medskip  \\
(0,3) & \quad &
{\textcolor{red}{(1,3)}} & \quad &
{\textcolor{blue}{(2,3)}} & \quad &
{\textcolor{green}{(3,3)}} & \quad &
{\textcolor{cyan}{[4,3]}} \medskip  \\
(0,2) & \quad &
(1,2) & \quad &
{\textcolor{red}{(2,2)}} & \quad &
{\textcolor{blue}{(3,2)}} & \quad &
{\textcolor{green}{[4,2]}} \medskip  \\
(0,1) & \quad &
(1,1) & \quad &
(2,1) & \quad &
{\textcolor{red}{(3,1)}} & \quad &
{\textcolor{blue}{[4,1]}} \medskip \\
(0,0) & \quad &
(1,0) & \quad &
(2,0) & \quad &
(3,0) & \quad &
{\textcolor{red}{[4,0]}}  \medskip
\end{tabular}
\end{center}
\end{table}

In our LB model (of order $N=3$), the Cartesian components of the
discrete velocity vectors are roots of the Hermite polynomial of
order $Q=N+1=4$.
Let us assume that at time $t=0$, the distribution function $f_{k}$,
$k = 1,\,2,\,\ldots K$, is
expressed as an expansion up to order $N=3$ with respect to the
tensor Hermite polynomials, Eq.~(\ref{exfh2}). 
According to Eq.~(\ref{lmn}), this means that $f_{k}$ contains
all the terms $H^{(m)}(\xi_{x})H^{(n)}(\xi_{y})$, with $0\leq m+n \leq 3$,
marked in black as $(m,n)$ in the lower left corner of Table \ref{t1}.
 After performing a time step $\delta t$, the expansion of
$f_k$ will include five new terms, namely the tensor Hermite polynomials of order $\ell=4$, in accordance to the recurrence relation \eqref{e3}.
 According to the recurrence relation for Hermite polynomials \eqref{e6},
these new terms of order 4 (marked in red color on
the {\emph{north-west -- south-east}} diagonal on Table \ref{t1})
are of the type $H^{(m)}(\xi_{x})H^{(n)}(\xi_{y})$, with $m+n = 4$.
Two of these terms, namely $H^{(4)}(\xi_{x})H^{(0)}(\xi_{y})$ and
$H^{(0)}(\xi_{x})H^{(4)}(\xi_{y})$, vanish because the components of
the velocity vectors used in this models are roots of the Hermite
polynomials of order N=4. The indices $m,n$ corresponding to
 these particular "red color" terms are evidenced by square brackets
(i.e., $[4,0]$ and $[0,4])$ in Table \ref{t1}.
At the next time step, the remaining (non-vanishing) red terms of
order $\ell=m+n=4$ evolve
further and produce the green terms in the table. Subsequent time steps
produce the terms marked cyan and magenta. Since $H^{(m)}(\xi_{x})H^{(n)}(\xi_{y}) = 0$ for
$m=4$ or $n=4$, the subsequent time steps never produce
non-vanishing terms of order $m+n>2N$ in the expression of $f_{k}$.


\newpage

\newpage

\newpage

\newpage

\newpage

\newpage

\newpage

\newpage

\begin{thebibliography}{99}


\bibitem{rallison-1984}
J. M. Rallison,
Annu. Rev. Fluid Mech. {\bf 16}, 45 (1984).

\bibitem{brennen-1995}
C. E. Brennen, {\it Cavitation and Bubble Dynamics}, (Oxford University, 
New York, 1995).

\bibitem{plesset-1963}
M. Plesset,
J. Fluid Eng. {\bf 85}, 360 (1963).

\bibitem{devega-2001}
H. J. de Vega, I. M. Khalatnikov, and N. G. Sanchez, eds.,
{\it Phase Transitions in the Early Universe: Theory and Observations}, 
(Springer, Berlin, 2001).

\bibitem{massol-2005}
H. Massol and T. Koyaguchi,
J. Volcanol. Geotherm. Res. {\bf 143}, 69 (2005).

\bibitem{kuksin-2010}
A. Y. Kuksin, G. E. Norman, V. V. Pisarev, V. V. Stegailov,
and A. V. Yanilkin, 
Phys. Rev. B {\bf{82}}, 174101 (2010).

\bibitem{watanabe-2010}
H. Watanabe, M. Suzuki, and N. Ito,
Phys. Rev. E {\bf{82}}, 051604 (2010).

\bibitem{diemand-2014}
J. Diemand, R. Angelil, K. K. Tanaka, and H. Tanaka, 
Phys. Rev. E {\bf{90}}, 052407 (2014).

\bibitem{angelil-2014}
R. Angelil, J. Diemand, K. K. Tanaka, and H. Tanaka, 
Phys. Rev. E {\bf{90}}, 063301 (2014).

\bibitem{sukop-2005}
M. C. Sukop and D. Or,
Phys. Rev. E {\bf{71}}, 046703 (2005).

\bibitem{chen-2010}
X.-P. Chen,
Commun. Comput. Phys. {\bf 7}, 212 (2010).

\bibitem{chen-2011}
X.-P. Chen, C.-W. Zhong, and X.-L. Yuan,
Comput. Math. Appl. {\bf 61}, 3577 (2011).

\bibitem{zhong-2012}
M. Zhong, C. Zhong, and C. Bai,
Adv. Comput. Sci. Appls {\bf 1}, 73 (2012).

\bibitem{feng-1997}
For a review, see for example,
Z. C. Feng and L. G. Leal,
Annu. Rev. Fluid Mech. {\bf 29}, 201 (1997).

\bibitem{rayleigh-1917}
L. Rayleigh,
Philos. Mag. {\bf 34} (1917).

\bibitem{plesset-1949}
M. Plesset,
J. Appl. Mech. {\bf 16}, 277 (1949).

\bibitem{plesset-1977}
M. S. Plesset and A. Prosperetti,
Annu. Rev. Fluid Mech. {\bf 9}, 145 (1977);
A. Prosperetti and M. S. Plesset,
J. Fluid Mech. {\bf 85}, 349 (1978).


\bibitem{chen_annurev1998}
S. Chen, G.D. Doolen,
Annu. Rev. Fluid. Mech. {\bf{30}} (1998) 329.

\bibitem{succi_2001}
S. Succi,
 {\it The Lattice Boltzmann Equation for Fluid Dynamics and Beyond},
(Clarendon Press, Oxford, 2001).

\bibitem{sukop_2006}
M.C. Sukop, D.T. Thorne,
 {\it Lattice Boltzmann Modeling: An Introduction for Geoscientists and
Engineers}, (Springer, Berlin, 2006).

\bibitem{aidun_annurev2010}
C. K. Aidun and J. R. Clausen,
Annu. Rev. Fluid. Mech. {\bf{42}} (2010) 439.

\bibitem{guo2013}
Z. Guo, C. Shu,
 {\it Lattice Boltzmann Method and its Applications in Engineering},
(World Scientific, Singapore, 2013).

\bibitem{sukop2015}
H.B. Huang, M.C. Sukop, X.Y. Lu,
Multiphase Lattice Boltzmann Methods: Theory and Application,
Wiley Blackwell, Chichester, 2015

\bibitem{krueger2017}
T. Kr\"{u}ger, H. Kusumaatmaja, A. Kuzmin, O. Shardt, G. Silva, E.M. Viggen,
 {\it The Lattice Boltzmann Method Principles and Practice},
(Springer, London, 2017).

\bibitem{shan_jfm2006}
X. Shan, X. Yuan, H. Chen,
J. Fluid. Mech {\bf{550}} (2006) 413.

\bibitem{cao1997}
N. Cao, S. Chen, S. Jin, D. Martinez,
Phys. Rev. E {\bf{55}} (1997) R21. 

\bibitem{dsfd2014}
T. Biciu\c{s}c\u{a}, A. Horga, V. Sofonea,
Comptes Rendus M\'{e}canique {\bf 343}, 580 (2015).

\bibitem{yuan2006}
P. Yuan and L. Schaefer, Phys. Fluids {\bf 18}, 042101 (2006).

\bibitem{chen2014}
For a review see, e. g., 
L. Chen, Q. Kang, Y. Mu, Y.-L. He, and W.-Q. Tao, Int. J. 
Heat and Mass Transfer {\bf 76}, 210 (2014).

\bibitem{kaehler-2015}
G. K\"{a}hler, F. Bonelli, G. Gonnella, and A. Lamura,
Phys. Fluids {\bf 27}, 123307 (2015).

\bibitem{rich1968}
S. Richardson, J. Fluid Mech.
{\bf 33}, 476 (1968).

\bibitem{buck1973}
J. D. Buckmaster and J. E. Flaherty, J. Fluid Mech.
{\bf 60}, 625 (1973).

\bibitem{hall1996}
I. Halliday and C. M. Care, Phys. Rev. E
{\bf 53}, 1602 (1996).

\bibitem{hall21996}
I. Halliday, C. M. Care, S. Thompson, and D. White, 
Phys. Rev. E
{\bf 54}, 2573 (1996).

\bibitem{wagner1997}
A. J. Wagner and J. M. Yeomans,
Int. J. Mod. Phys. C
{\bf 8}, 773 (1997).


\bibitem{colella1990}
P. Colella,
J. Comput. Phys. {\bf{87}} (1990) 171.

\bibitem{leveque1996}
R.J. Leveque,
SIAM J. Numer. Anal. {\bf{33}}, 627 (1996).

\bibitem{leveque2002}
R.J. Leveque,
 {\it Finite Volume Methods for Hyperbolic Problems},
(Cambridge University Press, Cambridge, 2001).

\bibitem{trangenstein2009}
J.A. Trangenstein,
 {\it Numerical Solution of Hyperbolic Partial Differential Equations},
(Cambridge University Press, Cambridge, 2009).

\bibitem{laurila-2012}
T. Laurila, A. Carlson, M. Do-Quang, T. Ala-Nissila, and G. Amberg,
Phys. Rev. E {\bf 85}, 026320 (2012).

\bibitem{falcucci-2013}
G. Falcucci, E. Jannelli, S. Ubertini, and S. Succi
J. Fluid Mech. {\bf 728}, 362 (2013).

\bibitem{pre2004}
V. Sofonea, A. Lamura, G. Gonnella, and A. Cristea,
Phys. Rev. E {\bf{70}} (2004) 046702;
A. Cristea, G. Gonnella, A. Lamura, and V. Sofonea,
Math. Comput. Simulat. {\bf 72}, 113 (2006).

\bibitem{ambrus_pre2012}
V.E. Ambru{\c{s}} and V. Sofonea,
Phys. Rev. E {\bf{86}} (2012) 016708.

\bibitem{ambrus_jcph2016}
V.E. Ambru{\c{s}} and V. Sofonea,
J. Comput. Phys. {\bf{316}} (2016) 760.

\bibitem{shan_prl1998}
X. Shan and X. He,
Phys. Rev. Lett. {\bf{80}} (1998) 65.

\bibitem{piaud_ijmpc2014}
B. Piaud, S. Blanco, R. Fournier, V.E. Ambrus, and V. Sofonea,
Int. J. Mod. Phys. C {\bf{25}} (2014) 1340016.

\bibitem{hildebrandt}
F.B. Hildebrandt, 
 {\it Introduction to Numerical Analysis (second edition)},
(Dover Publications, 1987).

\bibitem{shizgal}
B. Shizgal, 
 {\it Spectral Methods in Chemistry and Physics: Applications to
Kinetic Theory and Quantum Mechanics}, (Springer, 2015).

\bibitem{shan2008}
X. Shan, Phys. Rev. {\bf{77}} (2008) 066702.

\bibitem{suga2013fdr}
K. Suga,
Fluid. Dyn. Res. {\bf{45}} (2013) 034501.

\bibitem{niu2007pre}
X. Niu, S. Hyodo, T. Munekata,
Phys. Rev. E {\bf{76}} (2007) 036711.

\bibitem{suga2010pre}
K. Suga, S. Takenaka, T. Ito, M. Kaneda, T. Kinjo, S. Hyodo,
Phys. Rev. E {\bf{82}} (2010) 016701.

\bibitem{ansumali_epl2003}
S. Ansumali, I.V. Karlin, H.C. \"{O}ttinger,
Europhys. Lett. {\bf{63}} (2003) 798.

\bibitem{bardow_epl2006}
A. Bardow, I.V. Karlin, A.A. Gusev,
Europhys. Lett. {\bf{75}} (2006) 434.

\bibitem{bardow_pre2008}
A. Bardow, I.V. Karlin, A.A. Gusev,
Phys. Rev. E {\bf{77}} (2008) 025701(R).

\bibitem{luo1998prl}
L.S. Luo, Phys. Rev. Lett.  {\bf{81}} (1998) 1618.

\bibitem{luo2000}
L.S. Luo, Phys. Rev. E {\bf{62}} (2000) 4982.

\bibitem{cicp2010}
A. Cristea, G. Gonnella, A. Lamura, and V. Sofonea,
Commun. Comput. Phys. {\bf{7}} (2010) 350.

\bibitem{coclite}
A. Coclite, G. Gonnella, and A. Lamura, 
Phys. Rev. E {\bf{89}}, 063303 (2014).

\bibitem{rowl}
J. S. Rowlinson and B. Widom,
{\it Molecular Theory of Capillarity},
(Clarendon Press, Oxford, 1982).

\bibitem{evans}
R. Evans,
Adv. Phys. {\bf 28}, 143 (1979).

\bibitem{noitermico}
G. Gonnella, A. Lamura, and V. Sofonea,
Phys. Rev. E {\bf 76}, 036703 (2007);
G. Gonnella, A. Lamura, and V. Sofonea,
Eur. Phys. J. - Spec. Top. {\bf 171}, 181 (2009). 

\bibitem{klimontovich}
Y.L. Klimontovich, 
 {\it Kinetic Theory of Nonideal Gases and
Nonideal Plasmas}, (Pergamon Press, Oxford, 1982).

\bibitem{leclaire}
S.Leclaire, M. El-Hachem, J.Y.Trepanier, and M.Reggio,
J. Sci. Comput. {\bf{59}} (2014) 545.

\bibitem{kart}
M. Patra and M.Karttunen,
Numer. Methods Partial Differ. Eqs. {\bf{22}} (2006) 936.

\bibitem{mattila2014}
K.K.Mattila, L.A.Hegele, and P.C.Philippi,
Sci. World J. {\bf{2014}} (2014) 142907.

\bibitem{siebert2014}
D. N. Siebert, P. C. Philippi, and K. K. Mattila,
Phys. Rev. E {\bf{90}} (2014) 053310.

\bibitem{farber}
R. Farber, 
 {\it CUDA Application Design and Development},
*Morgan Kaufmann, Waltham, MA, 2011).

\bibitem{cook}
S. Cook, 
 {\it CUDA Programming, A developer's Guide to Parallel
Computing with GPUs}, (Morgan Kaufmann, Waltham, MA, 2013).

\bibitem{professionalCUDA}
J. Cheng, M. Grossman, and T. McKercher,
Professional CUDA C Programming,
John Wiley and Sons, Inc., Indianapolis, IN, 2014.

\bibitem{cudaguide}
{\emph{CUDA C Programming Guide}}, http://docs.nvidia.com/cuda/pdf/CUDA\_C\_Programming\_Guide.pdf.

\bibitem{deville}
M.O. Deville and T.B. Gatski,
 {\it Mathematical Modeling for Complex Fluids and Flows},
(Springer, Berlin, 2012).

\bibitem{philippi_pre2006}
P.C. Philippi, L.A. Hegele Jr., L.O.E. dos Santos, and R. Surmas,
Phys. Rev. E {\bf{73}} (2006) 056702.

\bibitem{siebert_pre2008}
D.N. Siebert, L.A. Hegele Jr., and P.C. Philippi,
Phys. Rev. E {\bf{77}} (2008) 026707.

\bibitem{surmas_eurJph2009}
R. Surmas, C.E. Pico Ortiz, and P.C. Philippi,
Eur. Phys. J. Special Topics {\bf{171}} (2009) 81.

\bibitem{chika_pre2009}
S.S. Chikatamarla and I.V. Karlin,
Phys. Rev. E {\bf{79}} (2009) 046701.

\bibitem{ansumali_pre2008}
W.P. Yudistiawan, S. Ansumali, and I.V. Karlin,
Phys. Rev. E {\bf{78}} (2008) 016705.

\bibitem{ansumali_pre2010}
W.P. Yudistiawan, S.K. Kwak, D.V. Patil, and S. Ansumali,
Phys. Rev. E {\bf{82}} (2010) 046701.

\bibitem{islb1997}
X.Y. He,
Int. J. Mod. Phys. C {\bf{8}} (1997) 737.

\bibitem{islb2004}
X.D. Niu, C. Shu, Y.T. Chew, and T.G. Wang,
J. Stat. Phys. {\bf{117}} (2004) 665.

\bibitem{jcph2009}
V. Sofonea,
J. Comput. Phys. {\bf{228}} (2009) 6107.

\bibitem{ubertini_ccph2008}
S. Ubertini, S. Succi,
Commun. Comput. phys. {\bf{3}} (2008) 342.

\bibitem{guo_pre2003}
Z. Guo, T.S. Zhao, Phys. Rev. E {\bf{67}} (2003) 066709.

\bibitem{lee_jcph2001}
T. Lee, C.L. Lin,
J. Comput. Phys. {\bf{171}} (2001) 336.

\bibitem{lee_jcph2003}
T. Lee, C.L. Lin,
J. Comput. Phys. {\bf{185}} (2003) 445.

\bibitem{hejranfar_pre2015}
K. Hejranfar, E. Ezzatneshan,
Phys. Rev. E {\bf{92}} (2015) 053305.

\bibitem{australia2014}
D.M. Bond, W. Wheatley, M.N. Macrossan, and M. Goldsworthy,
J. Comput. Phys {\bf{259}} (2014) 175.

\bibitem{nannelli1992}
F. Nannelli and S. Succi,
J. Stat. Phys. {\bf{68}} (1992) 401.

\bibitem{hchen_pre1998}
H. Chen,
Phys. Rev. E {\bf{58}} (1998) 3955.

\bibitem{rzhang_pre2001}
R. Zhang, H. Chen, Y. Qian, and S. Chen,
Phys. Rev. E {\bf{63}} (2001) 056705.

\bibitem{mauro_pre2010}
M. Sbragaglia and K. Sugiyama,
Phys. Rev. E {\bf{82}} (2010) 046709.

\bibitem{jcph2003}
V. Sofonea, R. F. Sekerka, J. Comput. Phys. {\bf{184}} (2003) 422.

\bibitem{cejp2004}
A.Cristea, V.Sofonea, Cent. Eur. J. Phys. {\bf{2}}, 382 (2004).

\bibitem{zhang2006pre}
R. Zhang, X. Shan, H. Chen,
Phys. Rev. E {\bf{74}} (2006) 046703.

\bibitem{sescu2014}
A. Sescu, R. Hixon, J. Sci. Comput. {\bf{61}} (2014) 327.

\bibitem{sescu2015}
A. Sescu, Adv. Differ. Equ. {\bf{2015}} (2015) 9.

\bibitem{latt2006mcs}
J. Latt, B. Chopard, Math. Comput. Simulat. {\bf{72}} (2006) 165.

\bibitem{colosqui2009pof}
C. Colosqui, H. Chen, X. Shan, I. Staroselsky,
Phys. Fluids {\bf{21}} (2009) 013105.

\bibitem{colosqui2010pre}
C. Colosqui,
Phys. Rev. E {\bf{81}} (2010) 026702.

\bibitem{montessori2014pre}
A. Montessori, G. Falcucci, P. Prestininzi, M. La Rocca, S. Succi,

Phys. Rev. E {\bf{89}} (2015) 053317.
\bibitem{mattila_pof2017}
K. K. Mattila, P. C. Philippi, L. A. Hegele, Jr.,
Phys. Fluids {\bf{29}} (2017) 046103.

\bibitem{ijmpf}
P. Fede, V. Sofonea, R. Fournier, S. Blanco, O. Simonin,
G. Lepout\`{e}re, and V. Ambru\c{s},
Int. J. Multiphase Flow {\bf{76}} (2015) 187.

\bibitem{sspre2005}
V. Sofonea and R.F. Sekerka, 
Phys. Rev. E {\bf{71}} (2005) 066709.

\bibitem{pre2014}                                                              
V.E. Ambru\c{s} and V. Sofonea, Phys. Rev. E {\bf{89}} (2014) 041301(R).          

\bibitem{sone}
Y. Sone
{\it Molecular Gas Dynamics : Theory, Techniques and Applications}
(Birkh\"{a}user, Boston, 2007).

\bibitem{karniadakis}
G. Karniadakis, A. Be\c{s}kok, N. Aluru,
{\it Microflows and Nanoflows: Fundamentals and Simulation}
(Springer, Berlin, 2005).


\bibitem{acrivos-stone}
A. Acrivos, Ann. NY Acad. Sci. {\bf 404}, 1 (1983);
H. A. Stone, Annu. Rev. Fluid Mech. {\bf 26}, 65 (1994).

\bibitem{chen-2008} 
X. B. Nie, X. Shan, and H. Chen,
EPL {\bf 81}, 34005 (2008).

\bibitem{taylor-1932}
G. I. Taylor,
Proc. R. Soc. A {\bf 138}, 41 (1932).

\bibitem{barthes-1973}
D. Barth\`es-Biesel and A. Acrivos,
J. Fluid Mech. {\bf 61}, 1 (1973).

\bibitem{wagn07}
A. J. Wagner and C. M. Pooley,
Phys. Rev. E {\bf 76}, 045702(R) (2007).

\end{thebibliography}
\end{document}